\documentclass[iop,revtex4]{emulateapj}
\usepackage{color}
\usepackage{amsmath}
\usepackage{xcolor}
\usepackage{mathrsfs}
\usepackage{natbib}
\usepackage{gensymb}

\shortauthors{Bryan et al.}
\shorttitle{Extrasolar Obliquity}

\begin{document}

\title{Obliquity Constraints on an Extrasolar Planetary-Mass Companion}

\author{
Marta L. Bryan\altaffilmark{1}, Eugene Chiang\altaffilmark{1}, Brendan P. Bowler\altaffilmark{2}, Caroline V. Morley\altaffilmark{2}, Sarah Millholland\altaffilmark{3}, Sarah Blunt\altaffilmark{4}, Katelyn B. Ashok\altaffilmark{2}, Eric Nielsen\altaffilmark{5}, Henry Ngo\altaffilmark{6}, Dimitri Mawet\altaffilmark{4}, Heather A. Knutson\altaffilmark{4}
}

\altaffiltext{1}{Department of Astronomy, 501 Campbell Hall, University of California Berkeley, Berkeley, CA 94720-3411, USA}

\altaffiltext{2}{Department of Astronomy, The University of Texas at Austin, Austin, TX 78712, USA}

\altaffiltext{3}{Department of Astronomy, Yale University, 52 Hillhouse Ave, New Haven, CT 06511, USA}

\altaffiltext{4}{Cahill Center for Astronomy and Astrophysics, California Institute of Technology,
1200 East California Boulevard, MC 249-17, Pasadena, CA 91125, USA}

\altaffiltext{5}{Kavli Institute for Particle Astrophysics and Cosmology, Stanford University, 452 Lomita Mall, Stanford, CA 94305, USA}

\altaffiltext{6}{NRC Herzberg Astronomy and Astrophysics, 5071 West Saanich Road, BC V9E 2E7, Canada}

\begin{abstract} 
We place the first constraints on the obliquity of a planetary-mass companion (PMC) outside of the Solar System.  
Our target is the directly imaged 
system 2MASS J01225093--2439505 (2M0122), which consists of a 120 Myr 0.4 M$_{\odot}$ star hosting a 12--27 M$_{\rm J}$ companion at 50 AU.  
We constrain all three of the system's angular momentum vectors: how the companion spin axis, the stellar spin axis, and the orbit normal are inclined relative to our line of sight. To accomplish this, we measure projected rotation rates ($v\sin i$) for both the star and the companion using new near-infrared high-resolution spectra with NIRSPEC at Keck Observatory.  We combine these with a new stellar photometric rotation period from $\textit{TESS}$ and a published companion rotation period from $\textit{HST}$ to obtain spin axis inclinations for both objects.  We 
also fitted multiple epochs of astrometry, including a new observation with NIRC2/Keck, to measure 2M0122b's orbital inclination.  The three line-of-sight inclinations 
place limits on 
the true de-projected companion obliquity and stellar obliquity.
We find that while the 
stellar obliquity marginally prefers alignment, the 
companion obliquity tentatively 
favors misalignment.  We evaluate
possible origin scenarios. While collisions, secular spin-orbit resonances, and Kozai-Lidov oscillations 
are unlikely, 
formation by gravitational instability in a gravito-turbulent disk---the scenario favored for brown dwarf companions to stars---appears promising.

\keywords{planetary systems -- techniques: high-resolution spectroscopy, high-contrast imaging, photometry  -- methods: statistical }
\end{abstract}

\section{Introduction}
The obliquity of a planet 
reflects its formative and
subsequent dynamical history \citep[e.g.][]{Lissauer1993}. For the
terrestrial and ice giant planets in the
Solar System, spin rate and direction
are affected by 
accretion of planetesimals
and giant impacts \citep[e.g.][]{Lissauer1991,Schlichting2007,Reinhardt2019},
long-term gravitational forcing by other bodies
\citep[e.g.][]{Laskar1993,Touma1993}, and dissipative effects associated with atmospheric tides and the core-mantle boundary
\citep[e.g.][]{Dobrovolskis1980,Correia2006}. Gas giant spins may begin aligned with the
orbit normals of their parent circumstellar
gas disks (by conservation of vorticity),
but can be torqued out of alignment
by sweeping secular spin-orbit resonances
driven by orbital migration
\citep[e.g.][]{Ward2004}.
All of these processes may play out for
extrasolar planets
\citep[e.g.][]{Millholland2019,Auclair2017}.

A full specification of a planet-star system's
3D angular momentum architecture includes
the star's spin,
the planet's spin,
and the mutual orbit.
Radial velocity measurements of the 
Rossiter-McLaughlin effect in transiting exoplanet
systems can probe two of these and
constrain the stellar obliquity:
the angle between the stellar spin vector
and the orbit normal. Famously large stellar
obliquities in hot Jupiter systems
suggest, e.g., planet-planet
gravitational interactions
that can raise orbital inclinations dramatically
\citep[e.g.][]{Naoz2011,Dawson2018}. 

At higher companion masses, measurements of the projected
rotation speeds $v\sin i$ in very-low-mass (VLM) binaries with comparable spectral types showed that if their components have comparable true rotation rates, then spin axes in some systems must be mutually inclined, and by extension at least one spin axis must be inclined relative to the orbital plane \citep{Konopacky2012}. At the same time, a detailed study of the tight L-dwarf binary 2MASSW J0746425+200032AB 
found spin equator planes and the orbit plane to be aligned  \citep{Harding2013}.

As yet no measurement has been made of an
exoplanetary obliquity. How can we get at this,
or at least start to? If we can measure the planet's
radius $R$ (say its effective blackbody radius,
derived from its intrinsic luminosity
and effective temperature), its 
rotation period $P_{\rm rot}$ from photometry,
and its projected rotation speed $v \sin i$
from spectral line broadening, then
the inclination of the spin axis
relative to our line of sight can be calculated as
\begin{equation} \label{one}
i = \arcsin \left[ \frac{P_{\rm rot} \times (v \sin i)}{2\pi R} \right] \,.
\end{equation}
Directly imaged ``planetary-mass companions'' (PMCs), many of which have inferred masses straddling the deuterium burning limit, are
excellent targets for carrying out such a procedure.
\citet{Bowler2016} catalogs 25 wide-separation ($\gtrsim$ 1$"$) PMCs that are young
($\lesssim$ 100 Myr) and therefore relatively bright
at near-infrared (NIR) wavelengths. 
For now we label these objects PMCs and not ``planets'' or ``brown dwarfs'' in recognition of their {\it a priori} unknown formation mechanism, whether by core accretion on the one hand, or gravitational instability/turbulent fragmentation on the other. By the time this paper concludes, however, we will have  made a case for one and not the other.

Here we present the first constraints on a extrasolar PMC obliquity.  We measure line-of-sight inclinations of all three angular momentum vectors in the directly imaged system 2MASS J01225093--2439505 (hereafter 2M0122). The system consists of a 0.4 M$_{\odot}$ star orbited by an L3.5$\pm$1.0 PMC of mass 12-27 M$_{\rm J}$ at a projected separation of 50 AU \citep{Bowler2013,Hinkley2015}. As a member of the AB Dor association, 2M0122 has an age of $120\pm10$ Myr \citep{Bowler2013,Malo2013}.  

The rest of this paper is organized as follows. In Section 2 we describe our observations, including NIR high-resolution spectroscopy with NIRSPEC/Keck, high-contrast imaging with NIRC2/Keck, and photometry with $\textit{TESS}$.  In Section 3 we detail our measurements of the three line-of-sight inclinations for the companion and host star spins and the orbit normal, and establish what constraints we can put on the PMC obliquity and the stellar obliquity.  In Section 4 we discuss and evaluate origin scenarios for these obliquities.
We conclude in Section 5.

\section{Observations}

\subsection{Keck/NIRSPEC High-Resolution Spectroscopy}
On UT November 3 2017 we observed the host star 2M0122 and companion 2M0122b in $\textit{K}$ band (2.03 - 2.38um) using the high-resolution near-infrared spectrograph NIRSPEC at the Keck II telescope, which (pre-upgrade) had a resolution of $\sim$25,000 at the time of these observations \citep{McLean1998}.  We carried out the observations in adaptive optics (AO) mode using the 0.041$\times$2.26 arcsec slit, to minimize contaminating light from the star at the location of the companion (1.45" separation).  We obtained spectra for the star and companion separately.  We observed the m$_{\rm K}$ = 9.2 mag host star with a single ABBA nod using an exposure time of 60 seconds for each image.  Subsequently, we observed the m$_{\rm K}$ = 14.0 mag companion with eight AB nods using an exposure time of 600 seconds for the first three AB nods, and an exposure time of 900 seconds for the final five AB nods.  

\subsection{Keck/NIRC2 Adaptive Optics Imaging}

On UT June 18  2019 we observed the 2M0122 system in $\textit{Ks}$ band using the near-infrared imager NIRC2 (PI: Keith Matthews) on the Keck II telescope.  We used natural guide star AO imaging and the narrow camera setting to achieve better contrast and spatial resolution.  We read out the full 1024$\times$1024 pixel NIRC2 array and used a three-point dither pattern to avoid the NIRC2 quadrant with elevated noise.  We obtained four usable images each with an integration time of 60 seconds. We calibrate and remove artifacts using the dome flat fields and dark frames. 

\begin{figure*}[t]
\centering
\includegraphics[width=\textwidth]{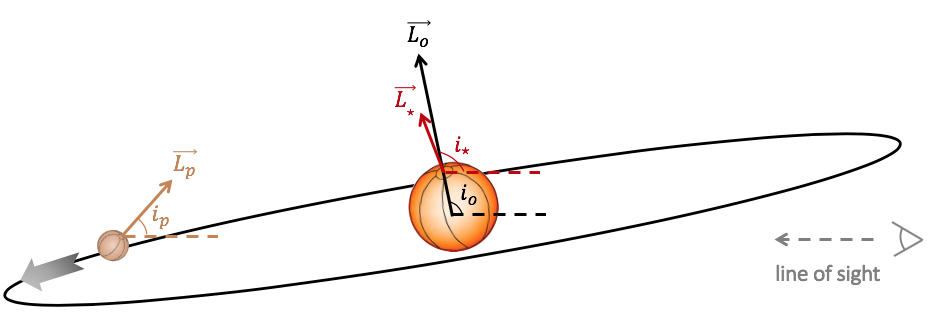}
\caption{The 3D geometry of the 2M0122 system 
is described
by three angular momentum vectors: $\vec{L}_o$ for the orbit, $\vec{L}_{\star}$ for the stellar spin,
and $\vec{L}_p$ for the PMC spin.
From our observations 
we measure these vectors' line-of-sight inclinations: $i_o$, $i_\star$, and $i_p$.
}
\label{fig:schematic}
\end{figure*}

\subsection{TESS Time-Series Photometry}

2M0122 was observed with the \emph{Transiting Exoplanet Survey Satellite} \citep{Ricker2015} in Sector 3 of the two-year primary mission between 2018 September 24 UT and 2018 October 14 UT (spacecraft orbits 13 and 14). This star was included in the \emph{TESS} input catalog (TIC ID 11614485; \citet{Stassun2018}) as part of the Cool Dwarf Sample \citep{Muirhead2018}. 2M0122 fell on CCD4 in Camera 2, and short-cadence science data with two-minute sampling were collected continuously for 20.4 days except during the central 1.1 days when the data were downlinked during perigee.

\section{Analysis}

Here we measure inclinations of all three angular momentum vectors in the 2M0122 system relative to the sky plane: the stellar spin angular momentum vector, the orbital angular momentum vector, and the PMC spin angular momentum vector.  These measurements yield 
the stellar spin axis inclination $i_{\star}$, the 
orbital inclination $i_{o}$,  and the PMC spin axis inclination $i_{p}$. (See Fig. \ref{fig:schematic} for a schematic representation.)  To obtain these inclinations, 
five quantities are needed:  the projected rotation rate $v_{\star} \sin i_{\star}$ for the star, the projected rotation rate $v_{p} \sin i_{p}$ for the companion, the rotation period $P_{rot,\star}$ for the star, the rotation period $P_{rot,p}$ for the companion, and
an astrometric orbit
for the companion.  We describe each of these five measurements below.

\subsection{Measuring $v \sin i$ for 2M0122 and 2M0122b}

Following the methodology outlined in \citet{Bryan2018}, we extract 1D spectra from our NIRSPEC images using a Python pipeline modeled after \citet{Boogert2002}.  We first flat-field, dark subtract, and then difference each AB pair. For each of the six orders, we then stack and align the differenced images to combine them into a single image. To correct the modest curvature of the 2D spectrum along the $x$ (dispersion) axis, we fit the spectral trace for each order with a third order polynomial.  We use the fit to the trace of the host star to rectify the 2D spectrum of both the star and the companion. This allows us to leverage the high signal-to-noise of the stellar trace and provide better constraints on the shape of the significantly fainter trace of the companion. While the star and the companion were not in the slit at the same time, the small 1.45" nod from the star to the companion did not significantly change the shape of the trace. 

We note that the pre-upgrade NIRSPEC detector occasionally produced one or more sets of every eight rows whose values were systematically off by an constant value.  This value was tied to one of the two quadrants on the left half of the detector.  These offsets are likely due to variations in bias voltages \citep{Bryan2018}.  While this effect is not significant when the signal-to-noise ratio of the trace is high, such as that of the host star 2M0122, it does become important when the traces are faint, as is the case for the companion 2M0122b.  For images of 2M0122b, we correct for this effect by calculating the median value of unaffected rows and subsequently adding or subtracting a constant value from the bad rows to match the median pixel value.  Unfortunately while this correction improved the noise in the left half of the detector, we find in subsequent analyses that including the left (blue) half of the companion spectrum degrades the significance of the cross correlation function and resulting spin measurement, which we detail later in this section. We thus do not consider the left (blue) half of the spectra for both companion and host star once their 1D spectra are extracted.

\begin{figure}[h]
\centering
\includegraphics[width=0.45\textwidth]{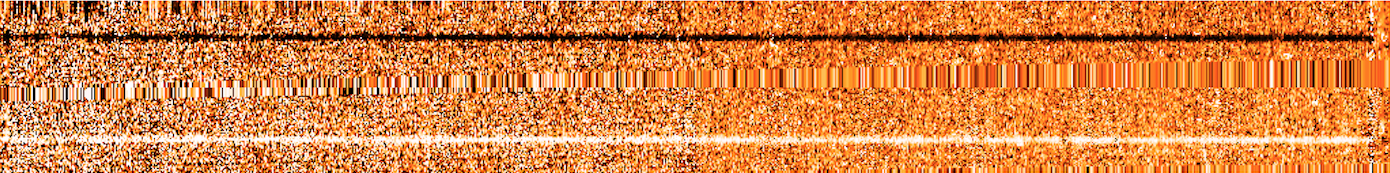}
\caption{2D rectified spectrum for 2M0122b order 2, wavelength range $2.27-2.31$ $\mu$m.  
Note that the left half of the order is noisier because of the imperfect correction of the every eighth row effect, which is likely due to bias voltage offsets.}
\label{fig:2D spectrum}
\end{figure}

1D spectra are then optimally extracted from 2D rectified spectra for each of the six orders (see Figure \ref{fig:2D spectrum} for an example 2D rectified order for 2M0122b).  For each positive and negative trace, we calculate an empirical PSF profile along the $y$ (cross-dispersion) axis of the 2D order.  We use this profile to combine the flux along each column, collapsing the 2D rectified order into a 1D spectrum in pixel space.  

After producing 1D spectra in pixel space for both the host star and companion, we wavelength calibrated each spectral order.  We first calculate the wavelength solution for the star by fitting the positions of the telluric lines in the spectrum with a third order polynomial wavelength solution $\lambda$ = $ax^3 + bx^2 + cx + d$, where $\lambda$ is wavelength and $x$ is pixel number.  While the 1D spectra for the companion have too low signal-to-noise ratios to confidently fit a third order polynomial wavelength solution, we note that because we maintained the same instrument configuration (filter, rotator angle, etc.) throughout the night, the wavelength solution should remain constant between the star and companion, aside from a linear offset due to the fact that the companion and star might have been placed at different positions in the slit.  To determine this linear offset, we apply the stellar wavelength solution to the companion spectra, and then cross correlate the 2M0122b spectrum with a telluric model.

\begin{figure}[h]
\centering
\includegraphics[width=0.5\textwidth]{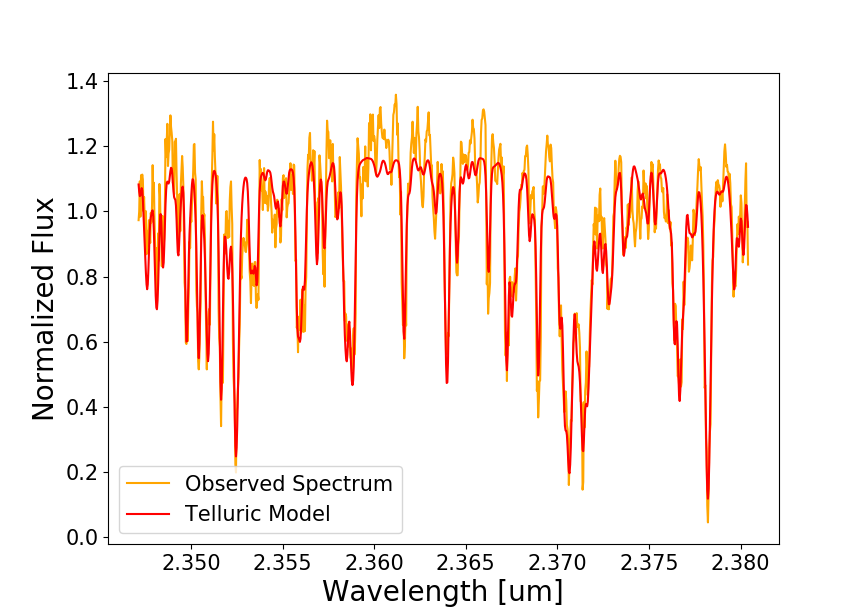}
\caption{Order 2 positive trace 1D wavelength calibrated spectrum of 2M0122 (orange), and the best fit \texttt{molecfit} telluric model (red). }
\label{fig:1D spectrum}
\end{figure}

To remove telluric features from the stellar spectrum, we use the \texttt{molecfit} routine, which simultaneously fits a telluric model and an instrumental profile defined by a single Gaussian kernel to the spectrum \citep{smette_molecfit2015, kausch_molecfit2015}.  In addition, \texttt{molecfit} iteratively fits the continuum with a third order polynomial before dividing out the telluric model.  We use the best fit telluric model for the host star 2M0122 to telluric correct the companion spectrum, dividing this model from the data.  We note that these telluric corrections leave significant artifacts in the spectra predominantly at the location of strong telluric lines, where the line cores are difficult to fit well.  We thus remove these artifacts at the locations of the strongest telluric absorption.  Figure \ref{fig:1D spectrum} shows an example 1D wavelength calibrated and telluric corrected spectrum for order 2 of 2M0122.

We note that for subsequent analyses we only use orders 1 and 2 (wavelength ranges $2.34-2.38$ $\mu$m and $2.27-2.31$ $\mu$m respectively) out of the full six orders for both the star and companion.  We utilize these two orders because they have the most accurate wavelength calibrations and telluric corrections, and they contain strong and numerous absorption lines from both water and CO, including two strong CO bandheads.  

With these 1D wavelength calibrated and telluric corrected spectra, we sought to measure rotational line broadening in those spectra, which yields the projected rotation rate $v\sin i$.  For both the star and companion, we measure $v\sin i$ and the radial velocity offset by calculating the cross-correlation function (CCF) between each observed spectrum and a model atmosphere, where the model atmosphere has been broadened to the instrumental resolution.  We use an atmospheric model from the Sonora model grid for the companion \citep[Marley et al. in prep., Morley et al. in prep.]{Marley2018}; these models are calculated assuming that the atmosphere is in radiative--convective and chemical equilibrium, following the approach of \citet{Marley1999, Saumon2008, Morley2012}, with updated chemistry and opacities as described in \citep[Marley et al., in prep.]{Marley2018}. We use a model with $T_{\rm eff}$ = 1600K and $log(g)$ = 4.5 for 2M0122b. These values were obtained from model fits to a low resolution SPHERE spectrum of 2M0122b \citep{Hinkley2015}. We assume solar metallicity and solar C/O ratio, and include silicate, iron, and corundum clouds with a sedimentation efficiency $f_{\rm sed}=2$ as described in \citet{Ackerman2001}. 

For the star, we use a BT Settl model with $T_{\rm eff}$ = 2500K and $log(g)$ = 5.0, where we determined these $T_{\rm eff}$ and $log(g)$ parameters from BT Settl isochrones corresponding to the measured age of the system, 120$\pm$10 Myr, and the measured bolometric luminosity of the star, $log(L_{bol}/L_{\odot})$ = $-1.72\pm0.11$ dex \citep{Bowler2013}. We note that the stellar mass corresponding to these $log(L_{bol}/L_{\odot})$ and age parameters is 0.4 $M_{\odot}$, which we use later in the analysis.

We then compare this ``data'' CCF to a series of ``model'' CCFs, where each ``model'' CCF was calculated by cross-correlating a model atmosphere broadened to the instrumental resolution, with that same model additionally broadened by a rotation rate and offset by a radial velocity (RV).  We carry out this comparison in a Bayesian framework using MCMC to fit for three free parameters:  $v\sin i$, RV, and instrumental resolution.  While we use uniform priors on $v\sin i$ and RV, we use a Gaussian prior for the instrumental resolution, with a peak location and width defined as 24,800 and 1000 respectively, in order to properly incorporate uncertainties on the measured instrumental resolution given the degeneracy between broadening due to instrumental resolution and rotational line broadening.  The measured instrumental resolution and corresponding uncertainty originate from a robust measurement of the instrumental resolution for NIRSPEC on the night that these observations were taken, from the same observing program and identical instrumental set-up, described in \citet{Xuan2020}.  In \citet{Xuan2020}, the authors used a previously published $v\sin i$ value for DH Tau to determine the broadening in the stellar spectrum due to the instrumental resolution, and confirmed that measurement matched their measurement of instrumental resolution from telluric line fits.  We also note that while we use a uniform prior on $v\sin i$ in this analysis, we test whether this choice could bias our resulting rotation rate measurement by instead separating $v$ and $i$ into two separate parameters that are varied in the MCMC, using a uniform prior on $v$ and a prior that is uniform in $\cos i$ for $i$.  We find that the resulting rotation rate is consistent with that determined just by varying $v\sin i$ with a uniform prior at the $<0.2 \sigma$ level.  

The log likelihood function used in our MCMC framework is given by
\begin{equation}
\log{L} = \sum_{i=1}^{n} -0.5\bigg(\frac{m_i - d_i}{\sigma_{i}}\bigg)^2,
\label{eq:loglike}
\end{equation}
\noindent where $d$ is the ``data'' CCF calculated by cross-correlating the observed spectrum with a model atmosphere broadened to the instrumental resolution, $m$ is the ``model'' CCF calculated by cross-correlating a model atmosphere broadened by the instrumental resolution with that same model additionally broadened by a $v\sin i$ value and offset by some radial velocity.  We calculate uncertainties on the ``data'' CCF using the jackknife resampling technique.  In this case, uncertainties are given by 
\begin{equation}
\sigma_{\rm{jackknife}}^2 = \frac{(n-1)}{n} \sum_{i=1}^{n} {(x_i - x)}^2,
\label{eq:jackknife}
\end{equation}
\noindent where $n$ is the total number of samples, here defined as the total number of AB pairs -- eight for 2M0122b and two for 2M0122 -- $x_i$ is the ``data'' CCF calculated using all AB pairs but the $i$th AB pair, and $x$ is the ``data'' CCF calculated using all AB pairs.

\begin{figure}[h]
\centering
\includegraphics[width=0.5\textwidth]{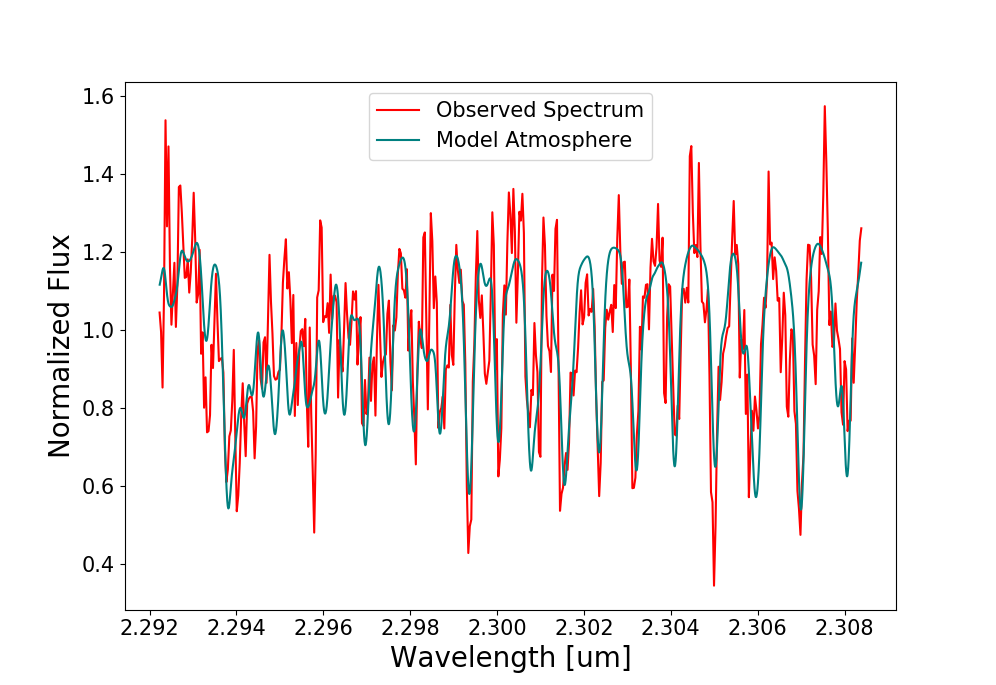}
\caption{Order 2 negative trace spectrum of 2M0122b (red).  Model atmosphere broadened by the instrumental resolution as well as the best-fit rotation rate, and shifted by the best-fit radial velocity offset (teal).  Note the strong CO bandhead at $\sim$2.294um.}
\label{fig:Spec/model comp}
\end{figure}

\begin{figure}[h]
\centering
\includegraphics[width=0.5\textwidth]{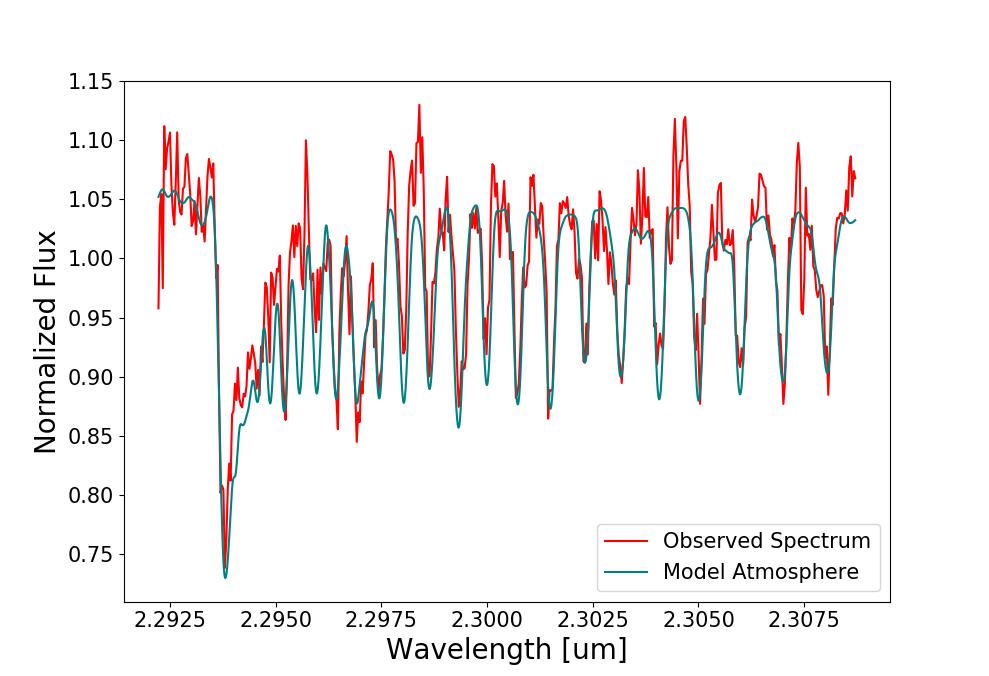}
\caption{Order 2 positive trace of 2M0122 (red). Model atmosphere broadened by the instrumental resolution as well as the best-fit rotation rate, and shifted by the best-fit radial velocity offset (teal).}
\label{fig:Spec/model star}
\end{figure}

\begin{figure}[h]
\centering
\includegraphics[width=0.5\textwidth]{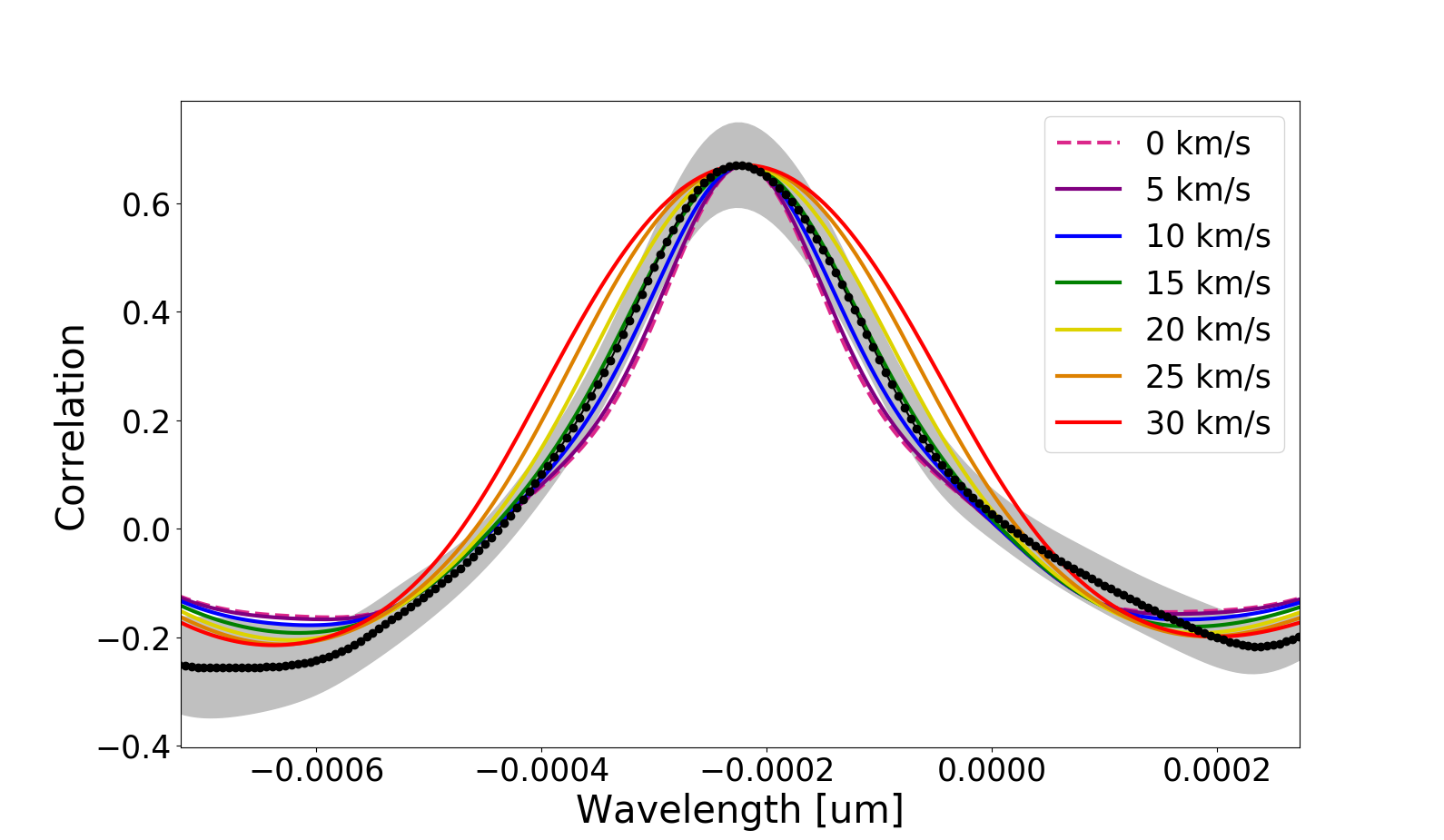}
\caption{Cross correlation function between the order 2 negative trace spectrum of 2M0122b with a model atmosphere broadened to the instrumental resolution (black points), shown with 1$\sigma$ uncertainties shaded in gray.  The cross-correlation functions between a model atmosphere broadened to the instrumental resolution, and that same model additionally broadened by a series of rotation rates (0, 5, 10, 15, 20, 25, 30 km/s) are shown in color. }
\label{fig:CCF negative 1}
\end{figure}

\begin{figure}[h]
\centering
\includegraphics[width=0.5\textwidth]{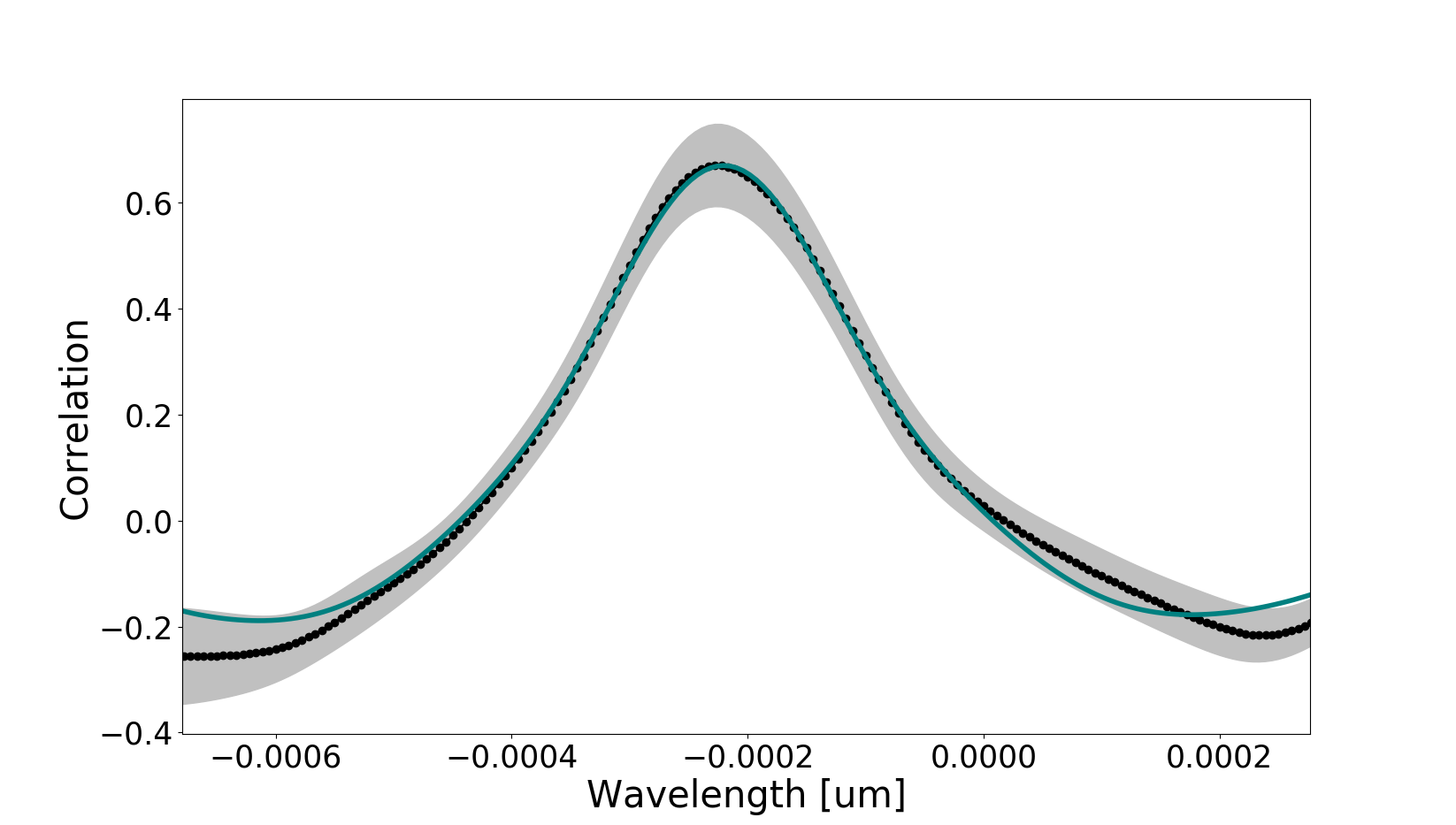}
\caption{Cross correlation function between the order 2 negative trace spectrum of 2M0122b with a model atmosphere broadened to the instrumental resolution (black points), shown with 1$\sigma$ uncertainties shaded in gray.  The cross-correlation function between a model atmosphere broadened to the instrumental resolution, and that same model additionally broadened by the best fit rotation rate and shifted by the best fit velocity offset are shown in teal. }
\label{fig:CCF negative 2}
\end{figure}

\begin{figure}[h]
\centering
\includegraphics[width=0.5\textwidth]{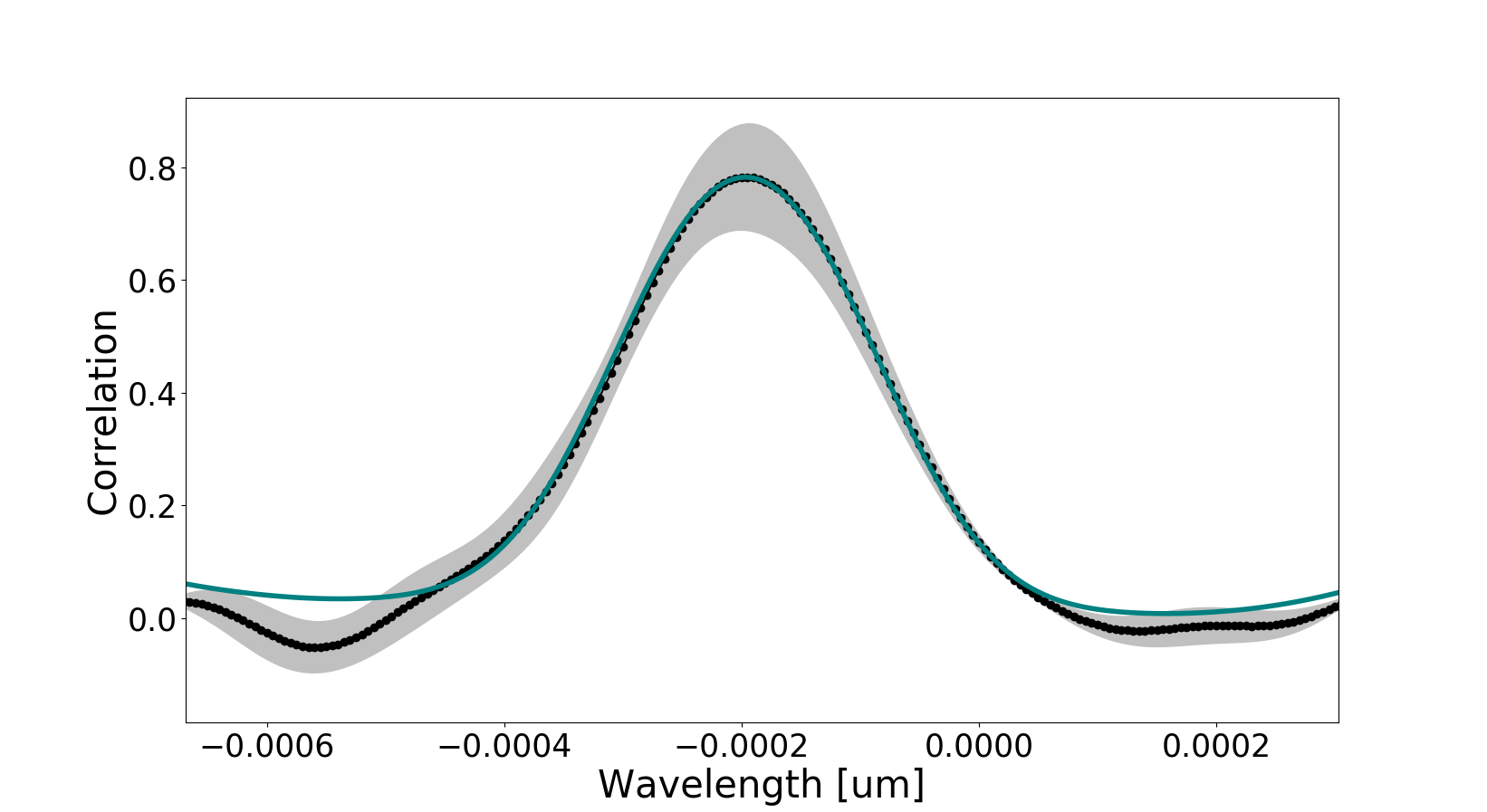}
\caption{Cross correlation function between the order 2 positive trace spectrum of 2M0122 with a model atmosphere broadened to the instrumental resolution (black points), shown with 1$\sigma$ uncertainties shaded in gray.  The cross-correlation function between a model atmosphere broadened to the instrumental resolution, and that same model additionally broadened by the best fit rotation rate and shifted by the best fit velocity offset are shown in teal. }
\label{fig:CCF positive}
\end{figure}

We next consider what assumptions on both the observational and the modeling side could impact measured rotation rates.  First, we investigate whether offsets in the relative positions of spectra from individual AB nod pairs could inflate the measured rotation rate.  To test this, we reduce each AB pair spectrum for 2M0122b and 2M0122 separately, and treated each positive and negative trace separately.  We compare the location of the CCF peaks in wavelength space.  We find that for the companion, CCF peak locations could differ by more than 10 km/s between AB pairs, and by more than 18 km/s between positive and negative traces.  Similarly, for the star we find that CCF peak locations could differ by as much as 6 km/s, and by as much as 8 km/s between positive and negative traces.  Treating each positive and negative trace separately, we thus shift the wavelengths of each individual AB pair spectrum according to the measured CCF peak offsets.  We then combine these shifted individual AB pair spectra separately for each positive and negative trace prior to implementing these spectra in the MCMC framework.  We note that by treating the positive and negative traces separately, we get independent estimates of $v\sin i$.  In addition, we fit for $v\sin i$ separately for orders 1 and 2.  Unfortunately for 2M0122b, we found that the order 1 ``data'' CCF did not contain a significant peak due to the low SNR of the order 1 spectrum with which we could measure a robust rotation rate. Thus for the companion we only fit the positive and negative traces of order 2, whereas for the star we fit positive and negative traces for both orders 1 and 2.  We find that the measured values are consistent within the uncertainties, and compute their error-weighted averages.  For 2M0122b, the measured projected rotation rate $v\sin i$ = $13.4^{+1.4}_{-1.2}$ km/s (Figures \ref{fig:Spec/model comp}, \ref{fig:CCF negative 1}, and \ref{fig:CCF negative 2}).  For 2M0122, $v\sin i$ = $18.2^{+0.5}_{-0.4}$ km/s (Figures \ref{fig:Spec/model star} and \ref{fig:CCF positive}).

To test our assumptions on the modeling side, we first investigate our choice of $T_{\rm eff}$ and $\log(g)$ that we assume to generate the models.  We take the published values $T_{\rm eff}$ = 1600 $\pm$ 100K and $\log(g)$ = 4.5$\pm$0.5 for 2M0122b from \citet{Hinkley2015}, and compute models with the following four combinations of parameters:  (1700K, 4.0dex), (1700K, 5.0dex), (1500K, 4.0dex), (1500K, 5.0dex).  To test the impact of the uncertainties on $T_{\rm eff}$ and $\log(g)$ on our measured rotation rate, we rerun our MCMC fits to the companion spectra using models calculated with these alternate sets of parameters.  We find that in all cases, the measured $v\sin i$ values from these new fits differ from the original $v\sin i$ value by less than 0.7$\sigma$.

We next test whether variations in the atmospheric C/O ratio adopted in the model used to measure rotation rates could impact the final $v\sin i$ measurement.  We generate models with the following C/O ratios:  0.25$\times$solar, 0.5$\times$solar, and 1.5$\times$solar, where the original model we used has a solar (0.54) C/O ratio.  We repeat our MCMC analysis to calculate new projected rotation rates with each of these models, and find that the resulting $v\sin i$ values differ from the original by less than 0.8$\sigma$.  

Finally, we test the effect of pressure broadening uncertainties on derived rotation rates by running additional models with modified molecular opacities. Two models were run using molecular cross sections that were 10$\times$ and 0.1$\times$ the actual pressure for the whole profile, simulating a scenario where the pressure broadening parameters used to create the cross sections are wrong by an order of magnitude. Another model used cross sections with minimal pressure broadening, assuming P=10$^{-6}$ bar for the molecular cross sections for the whole profile, to determine how including the pressure broadened cross sections affects the derived rotation rate for this object. Collision-induced opacity of hydrogen and helium was treated separately for all models, using the standard pressure for each layer. We find that the resulting $v\sin i$ values differ from the original by less than 0.6$\sigma$.

\subsection{Measuring P$_{rot,\star}$ for 2M0122}
\begin{figure*}[h]
\centering
\includegraphics[width=\textwidth]{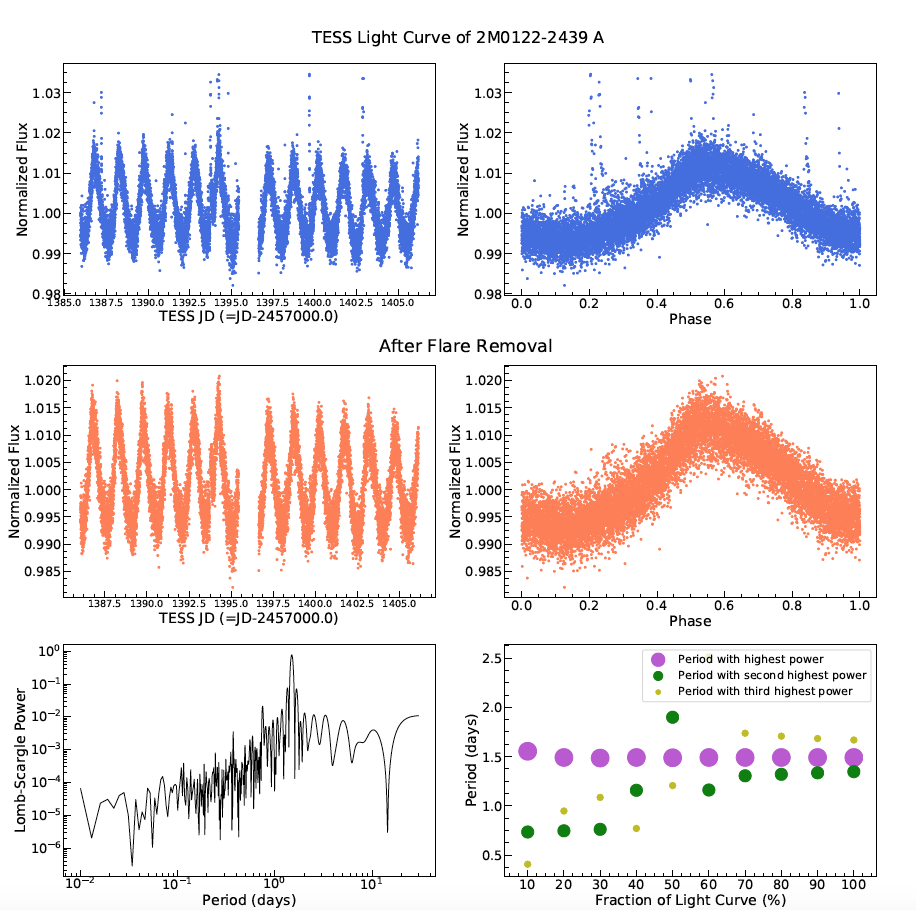} 
\caption{\emph{TESS} light curve of 2M0122 A.  Upper panels: extracted light curve (left), and the same data phase folded to the highest Lomb-Scargle periodogram power (right).  Several flares are visible in the data.  Middle panels: Light curve of 2M0122 A removing flares through the outlier rejection process described in the main text.  Our final period  of 1.49 d is derived from the flare-free light curve using a Lomb-Scargle periodogram (bottom right).  The phase-folded diagram shows little differential spot evolution.  Following \citet{Bowler2017}, we test whether smaller fractional light curve coverage influences the inferred period (bottom right); the peak periodogam power returns similar periods after iteratively removing up to 80\% of the data, implying the inferred period is robust against the timeframe of the observations.}
\label{fig:stellar_rotation}
\end{figure*}

We downloaded the high-cadence light curve using the \texttt{lightkurve} v.1.0.1 Python package (Lightkurve Collaboration, 2018), which interfaces with the \emph{TESS} data archive at the Mikulski Archive for Space Telescopes (MAST). Data with quality flags indicating potential anomalies have been removed, leaving 13,520 photometric points. The light curve of 2M0122 exhibits periodic modulations with a peak-to-peak amplitude of about 3\%.   12 full period modulations are evident, and about a dozen flares are visible  with a characteristic rise and exponential decay (Fig. \ref{fig:stellar_rotation}). Flares are removed by first identifying $>$5$\sigma$ local positive outliers in the light curve and then removing deviates in the residuals of a boxcar-smoothed and subtracted light curve. A Lomb-Scargle periodogram \citep{Lomb1976,Scargle1982} is used to assess periodic signal in the normalized (flare-free) light curve. The strongest power (reaching 0.735) is at 1.492 d, with the next highest peak at 1.348 days (reaching a power of 0.073) . To determine whether this periodic signal depends on the number of period cycles sampled, we repeated this experiment sequentially for progressively smaller portions of the light curve  following \citet{Bowler2017}.  The primary signal remained near 1.5 d even using only 10\% of the data, implying the measured period is robust against changes to the observational baseline. The phase-folded light curve does not show significant spot evolution over the 20-day timescale of the observations (Fig. \ref{fig:stellar_rotation}).

Differential rotation complicates the interpretation of rotationally modulated light curves because the latitude of the dominant starspots can bias the disk-integrated signal to shorter or longer periods.  We adopt a similar approach to that taken in \citet{Bowler2017} to estimate the uncertainty in the rotation period based on measurements of the pole-to-equatorial shear ($\Delta$$\Omega$) from \citet{Reinhold2015}, where  $\Delta$$\Omega$ = 2$\pi$(1/$P_\mathrm{min}$ -- 1/$P_\mathrm{max}$). If we assume the dominant spots are located at intermediate latitudes, we can estimate the uncertainty in the period ($\sigma_P$) as $P_\mathrm{min}$ = $P_\mathrm{measured}$ -- $\sigma_P$ and $P_\mathrm{max}$ =  $P_\mathrm{measured}$ + $\sigma_P$. Solving for $\sigma_P$ and adopting the maximum shear value of $\Delta$$\Omega$ $\approx$ 0.1 rad day$^{-1}$ found by \citet{Reinhold2015} for M dwarfs gives $\sigma_P$ = 0.02 d. Our final measurement for the rotation period of 2M0122 is 1.49 $\pm$ 0.02 d.

\subsection{Measuring P$_{rot,p}$ for 2M0122b}

The photometric rotation period for 2M0122b was published by \citet{Zhou2019}.  In this paper, the authors use $\textit{HST}$ Wide Field Camera 3 near-IR time-resolved photometry to measure photometric modulations in the light curve of 2M0122b.  These photometric modulations could come from either cloud patchiness or variability introduced by longitudinal bands, analogous to those seen on Jupiter.  By implementing techniques such as two-roll differential imaging and hybrid point-spread function modeling, the authors achieve a sub-percent photometric precision for this observation.  Using a Lomb-Scargle periodogram, the authors determine that the rotation period for 2M0122b is 6.0$^{+2.6}_{-1.0}$ hours.  Fitting the light curve with a sinusoid of period 6.0 hours, the best fit modulation amplitude is 0.52$\%$ $\pm$ 0.11$\%$.  

The authors note several caveats to the rotation rate measurement.  First, the 6.0 hour period detection has a significance of only 2.7$\sigma$.  In addition, given the low significance of the detection, it is not possible to constrain whether the companion light curve deviates from a single sinusoid, or whether it could have multiple peaks.  A previous study by \citet{Apai2017} found that for high S/N spectra with extremely long baselines ($>1$ year) for 3 L/T transition brown dwarfs, the power spectra of their light curves produced peaks at both the full rotation period of the object as well as half the rotation period.  However, even though the 2M0122b light curve has low S/N, both theoretical and higher quality light curves of brown dwarfs show that the full rotation periods are the dominant signal in the power spectrum of a given light curve \citep{Apai2017,Zhang2014}.  In addition, observations of Neptune and Jupiter show that the periodicity in their lightcurves corresponds to their full rotation periods \citep{Karalidi2015,Simon2016,Ge2019}.  In this paper, we thus proceed with the assumption that the 6.0 hour measured rotation period reflects the full rotation period.  

\subsection{Measuring astrometry for 2M0122b}
For each NIRC2 image of 2M0122b, we apply the NIRC2 detector dewarping solution~\citep{Service2016} and then fit for the pixel position of each source. We do this with a simultaneous fit on both sources' point spread functions using a combined Moffat and Gaussian functional form as described in~\citet{Ngo2015}. We then use the \citet{Service2016} plate scale ($9.971 \pm 0.004 \pm 0.001$ mas per pixel) and the North-alignment derotation correction (see footnote 13 in \citet{Bowler2018}) to compute the separation and position angle of the companion in each frame, relative to the primary star. Finally, we report the median of all four frames and the standard error on the median as the uncertainty. We find that the separation is $1451.3 \pm 2.9 \pm 1.6$ mas (total uncertainty: 3.3 mas) and the position angle is $215.44$ deg $\pm 0.12 \pm 0.06$ deg (total uncertainty 0.13 deg). For both values, the reported uncertainties are measurement uncertainty and the \citet{Service2016} NIRC2 distortion solution uncertainty, respectively.

\subsection{Measuring i$_{o}$}
\label{sec:measuring i_orbit}

We use the open-source Python package \texttt{orbitize!}\footnote{https://github.com/sblunt/orbitize} \citep{Blunt2019} to perform an orbit fit to the assembled astrometric measurements (summarized in Table \ref{tb:astrom}). We ran the \texttt{orbitize!} implementation of the Orbits for the Impatient (OFTI) algorithm \citep{Blunt2017} until $10^5$ orbits had been accepted. Briefly, OFTI is a modified rejection-sampling algorithm that generates independent sets of orbital parameters from a posterior distribution. It is more efficient than many MCMC implementations for orbits with small fractional orbit coverage. 

We parameterized the Keplerian orbit as: semimajor axis ($a$), eccentricity, inclination angle, argument of periastron ($\omega$), position angle of nodes ($\Omega$), and time of periastron passage (expressed as a fraction of the orbital period past MJD=58849; $\tau$). Parallax ($\pi$) and total mass ($M_T$) were also included as free parameters in the fit. We placed uniform priors on $\log{a}$, eccentricity, $\cos{i}$, $\omega$, $\Omega$, and $\tau$, and Gaussian priors on $\pi$ and $M_T$. Our prior on $\pi$ was constructed using the Gaia DR2 parallax and uncertainty for 2M0122 \citep{Gaia2018}, and our total mass prior was taken to be a Gaussian with $\mu=0.41M_{\odot}$ and $\sigma=0.08M_{\odot}$ \citep{Bowler2013}. 

\begin{figure}[h]
\centering
\includegraphics[width=0.5\textwidth]{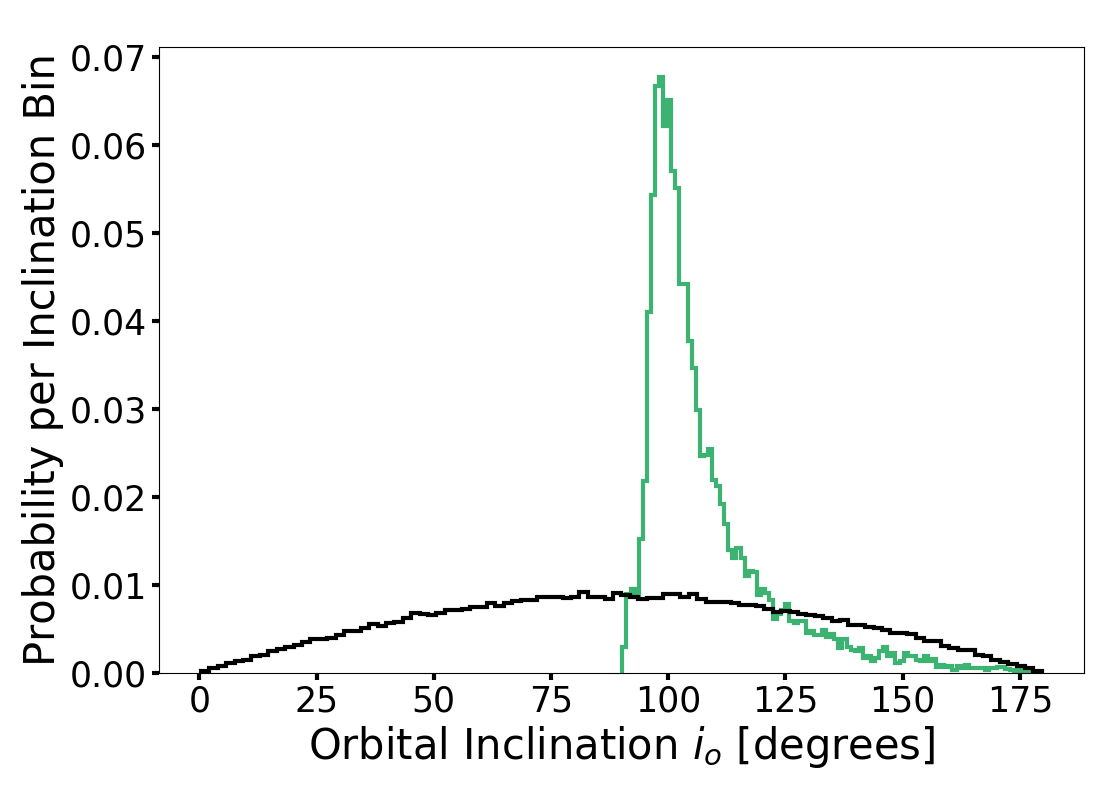} 
\caption{Normalized probability distribution of the orbital inclination (green).  The mode and 68$\%$ confidence interval of this angle are 103$^{+16}_{-6}$ degrees. This distribution is compared to a random inclination distribution (black) whose  values are drawn from a uniform distribution in $\cos i$.}
\label{fig: orbital inclination posterior}
\end{figure}

\begin{figure*}[]
    \centering
    \includegraphics[width=\textwidth]{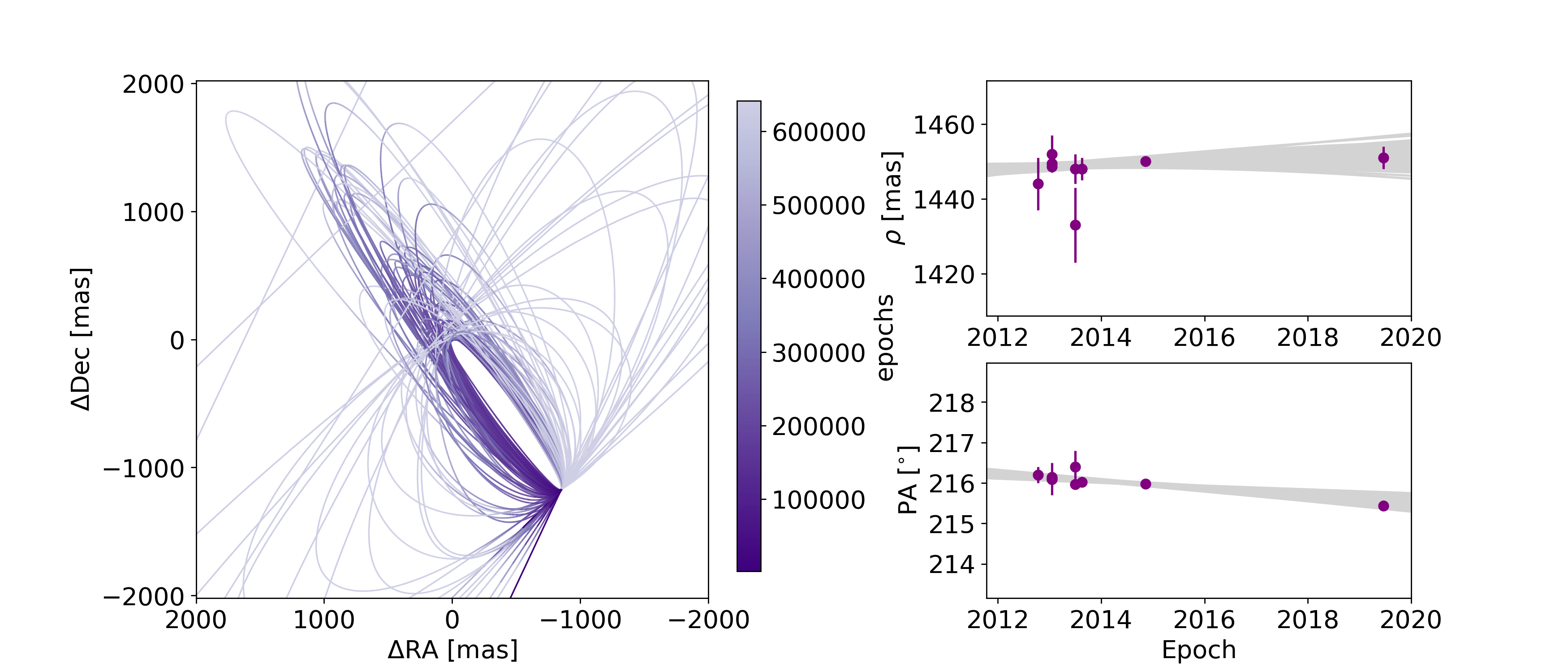}
    \caption{100 orbits of 2M0122b randomly drawn from the posterior. Left: orbits projected on the sky. The primary, 2M0122, is at (0,0). Orbits are color-coded by year, with darkest purple corresponding to the date of the first observation. Right top: separation versus time, with the posterior samples in grey and the data in dark purple. Right bottom: same for position angle. Linear orbital motion is apparent in separation and position angle.}
    \label{fig:orbit}
\end{figure*}

\begin{figure*}[]
    \includegraphics[width=8in]{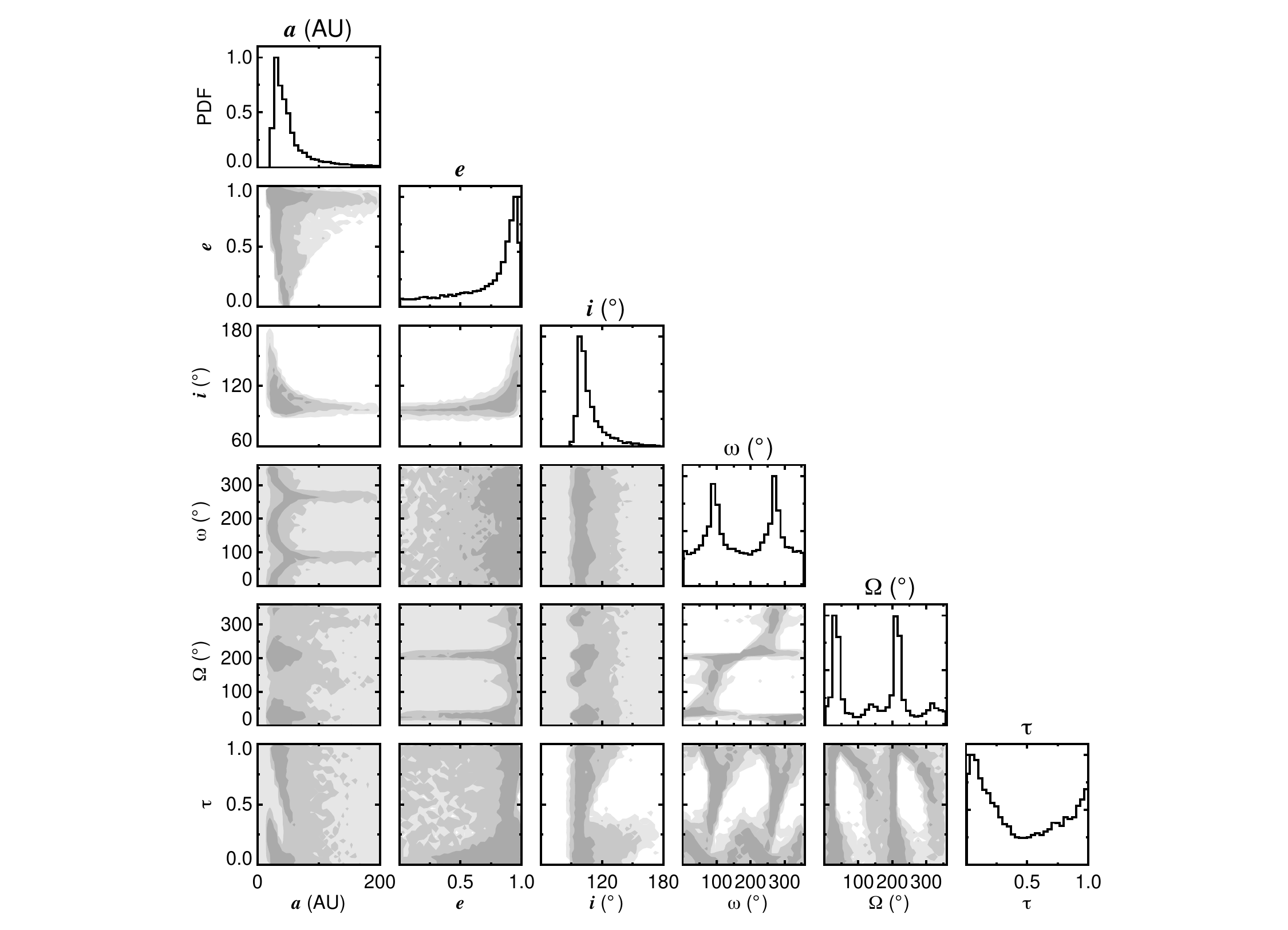}
    \caption{Corner plot showing marginalized one-dimensional posteriors (diagonal panels) and two-dimensional covariances (off-diagonal panels) between fitted orbital parameters. Eccentric, edge-on orbits are preferred.}
    \label{fig:corner}
\end{figure*}

2M0122b exhibits approximately linear orbital motion in separation ($\rho$) and position angle ($\theta$), which translates into a relatively narrow inclination angle constraint ($i=103^{+16}_{-6}$ deg). Figure \ref{fig: orbital inclination posterior} illustrates how this orbital inclination distribution compares to a random inclination distribution with values drawn from a uniform distribution in $\cos i$.  Eccentric, edge-on orbits are preferred. See Figures \ref{fig:orbit} and \ref{fig:corner} for visualizations of these orbits.

To bolster our confidence that the preference for high eccentricities is physical and not an artifact of the fitting process, we performed the following tests. First, we performed linear least-squares fits to separation and P.A. as functions of time, following Bowler et al (submitted). We generally expect linear evolution in separation and P.A. for the long orbits of directly-imaged objects, but if significant systematics are present in the data, we would expect a linear fit to yield a large $\chi^2_{\nu}$. We obtained $\chi^2_{\nu}<1$ for both the separation and P.A. fits, suggesting that unaccounted-for systematics are not driving the shape of the orbit. Second, we reran the orbit-fit with larger values of total mass (1.5 and 2.0x the actual mass). An underestimated total mass can bias the fit toward higher eccentricities, but we found that the preference for high eccentricities persisted when using larger values for the total mass. Furthermore, the orbital inclination was consistent within 1$\sigma$ for all three fits. Finally, we investigate the impact that a systematic offset on the final epoch of astrometry would have on the measured orbital inclination.  We increase the position angle of this final epoch by 0.4 degrees, a value consistent with the variations in PA amongst the earlier epochs taken within a short timespan, and rerun the orbit fit.  We find that the posterior on the inclination angle corresponds to 98$^{+12}_{-4}$ degrees, consistent with the original value of 103$^{+16}_{-6}$ degrees at $<$0.4$\sigma$.

\begin{deluxetable*}{cccccc}
\tablecaption{ \label{tb:astrom}}
\tabletypesize{\footnotesize}
\tablehead{
  \colhead{Epoch } & 
  \colhead{$\rho$ } & 
  \colhead{$\sigma_{\rho}$ } & 
  \colhead{$\theta$ } & 
  \colhead{$\sigma_{\theta}$ } & 
  \colhead{Reference}\\
  \colhead{[yr]} & 
  \colhead{[mas]} & 
  \colhead{[mas]} & 
  \colhead{[deg]} & 
  \colhead{[deg]} & 
  \colhead{}
}
\startdata
2012.780 & 1444 & 7 & 216.2 & 0.2 & \citealt{Bowler2013} \\
2013.047 & 1448.6 & 0.6 & 216.14 & 0.08 & \citealt{Bowler2013} \\
2013.047 & 1449.5 & 1.5 & 216.09 & 0.08 & \citealt{Bowler2013} \\
2013.049 & 1452 & 5 & 216.1  & 0.4 & \citealt{Bowler2013} \\
2013.493 & 1448 & 4 & 215.97 & 0.07 & \citealt{Bowler2013} \\
2013.493 & 1433 & 10 & 216.4 & 0.4 & \citealt{Bowler2013} \\
2013.626 & 1448 & 3 & 216.02 & 0.09 & \citealt{Bowler2015} \\
2014.858 & 1450 & 1 & 215.98 & 0.02 & \citealt{Bryan2016} \\
2019.460 & 1451 & 3 & 215.44 & 0.13 & This Work 
\enddata
\tablecomments{The listed PA values are 0.5\degree off from published values due to an error in the published North alignment correction  \citep{Bowler2018}.}
\end{deluxetable*}

\subsection{Measuring $i_{p}$ and $i_{\star}$}

From equation (\ref{one}), our measurements of rotation period $P_{\rm rot}$ and projected rotation speed $v\sin i$ can be combined to measure the line-of-sight spin axis inclination.
This can be done for both companion ($i_p$) and star ($i_{\star}$).  
For the radius $R$ in equation (\ref{one})
we employ the effective blackbody radius
\begin{equation} \label{radius}
R = \sqrt{\frac{L}{4\pi \sigma_b T_{\rm eff}^4}},
\end{equation}
\noindent where $L$ is the bolometric luminosity, $\sigma_b$ is the Stefan-Boltzmann constant, and $T_{\rm eff}$ is the effective temperature. Then
we re-write (\ref{one}) as

\begin{equation} \label{master}
i = \sin^{-1} \left[ \sqrt{\frac{\sigma_b}{\pi L}}  P_{\rm rot}T_{\rm eff}^2 \times (v \sin i) \right]
\end{equation}

\begin{figure}[h]
\centering
\includegraphics[width=0.5\textwidth]{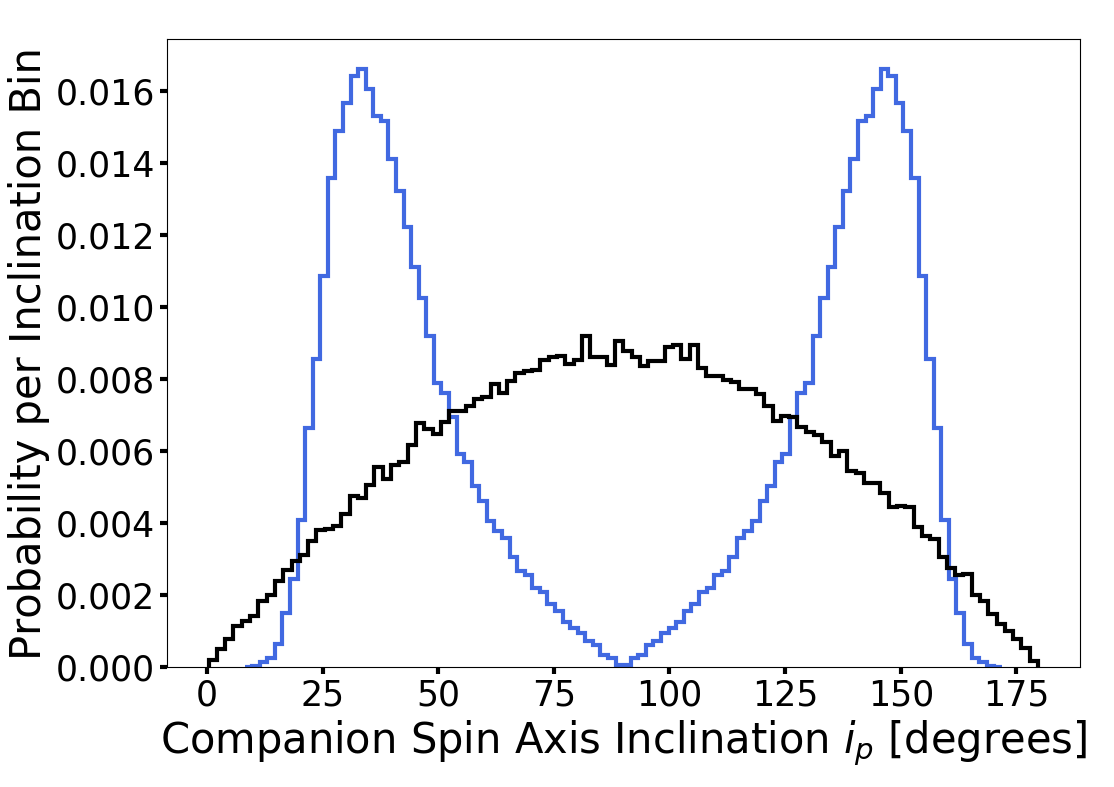}
\caption{Normalized probability distribution of the line-of-sight inclination of the companion spin axis (blue).
Putting aside the formally allowed values of $i_p > 90^\circ$, the mode and 68$\%$ confidence interval of $i_p$ are 33$^{+17}_{-9}$ degrees. This distribution is compared to a random inclination distribution (black) whose values are drawn from a uniform distribution in $\cos i$.}
\label{fig: planetary spin axis posterior}
\end{figure}

For 2M0122b, we adopt $\log(L_{bol}/L_{\odot})$ = $-4.19\pm0.10$ dex from \citet{Bowler2013} and $T_{\rm eff}$ = 1600$\pm$100K from model fits to a low-resolution (R$\sim$350) near-infrared spectrum taken with the SPHERE instrument on the VLT \citep{Hinkley2015}.  
From these measurements and equation (\ref{radius}),
we infer a companion radius of $R_p = 1.04 \pm 0.16 \,R_{\rm J}$.
For the full calculation of $i_p$ using (\ref{master}),
we propagate uncertainties using the Monte Carlo technique, drawing $10^6$ trials from normal distributions defined by the best fit values and uncertainties for each parameter.  We remove unphysical values for which $\sin i$ $>$ 1. 
Figure \ref{fig: planetary spin axis posterior} shows the resulting posterior distribution for $i_{p}$ in comparison to a random inclination distribution drawn from a uniform distribution in $\cos i$. 
While the random distribution is broad and peaks at $90$ degrees, the bimodal distribution for $i_p$ exhibits tighter constraints, favoring values near 33 and 147 degrees.
This distribution is symmetric about 90 degrees because we do not know whether the companion spin angular momentum vector is pointing towards us (at an angle of 33 degrees) or away from us (at an angle of 147 degrees). For $i_p < 90^\circ$,
the mode and 68$\%$ confidence interval are $i_p = 33^{+17}_{-9}$ degrees.  

We can make a quick consistency check on these results by noting
that $i_p = 90^\circ$ yields a lower bound
on the companion radius of $\min R_p = P_{\rm rot,p} \times (v_p \sin i_p) / (2\pi \sin i_p) \simeq 0.6 R_{\rm J}$.
This is consistently smaller than the value of
$R_p \simeq 1 \,R_{\rm J}$
computed using $L$ and $T_{\rm eff}$.

For the host star 2M0122, the measured bolometric luminosity is $\log(L_{bol}/L_{\odot}) = -1.72\pm0.11$ dex from \citet{Bowler2013}.  The effective temperature of the star is taken from \citet{Herczeg2014}, who derived empirical calibrations between spectral type and effective temperature for pre-main sequence stars.  Given the previously determined spectral type of M3.5$\pm$0.5 for 2M0122 \citep{Riaz2006}, we find that the spectral type falls between the M3.0 conversion to $T_{\rm eff}$ = 3410K, and the M4.0 conversion to $T_{\rm eff}$ = 3190K.  We take the midpoint of these two values $3300$K to be the effective temperature of 2M0122, with uncertainties 110K given the $\pm0.5$ dex uncertainty on the stellar spectral type.  

\begin{figure}[h]
\centering
\includegraphics[width=0.5\textwidth]{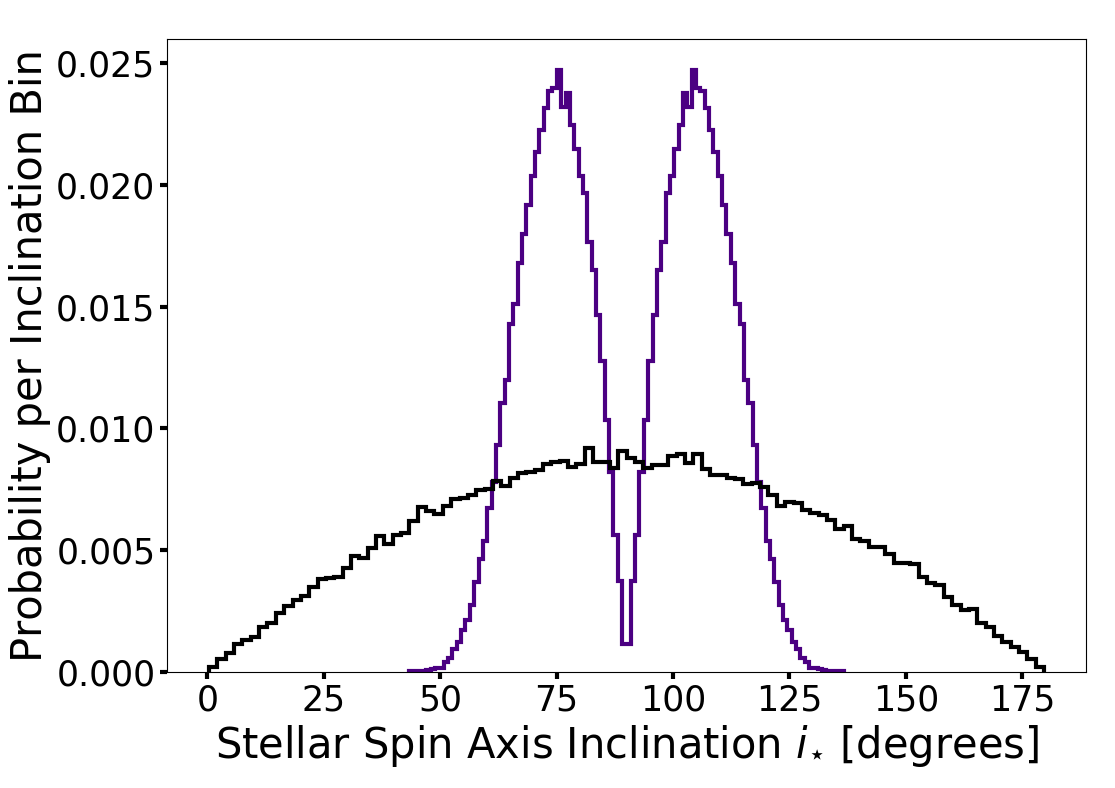}
\caption{Normalized posterior distribution of the line-of-sight inclination of the stellar spin axis. 
Putting aside the formally allowed values
of $i_\star > 90^\circ$, the mode and 68$\%$ confidence interval of $i_\star$ are $75 \pm 8$ degrees. This distribution is compared to a random inclination distribution (black) whose values are drawn from a uniform distribution in $\cos i$.}
\label{fig: stellar spin axis posterior}
\end{figure}

We propagate uncertainties using the Monte Carlo technique, drawing $10^6$ values from normal distributions defined by the best fit values and uncertainties on those values for each parameter.  We remove unphysical values for which $\sin i$ $>$ 1. Figure \ref{fig: stellar spin axis posterior} shows the resulting posterior distribution for $i_{\star}$ (again, symmetric about 90$^\circ$), and compares this to a random inclination distribution. We find that the mode and 68$\%$ confidence interval are $i_\star = 75 \pm 8$ degrees.

\subsection{Measuring the 3D spin-orbit architecture of the 2M0122 system} \label{3D}

\begin{figure}[h]
\centering
\includegraphics[width=0.5\textwidth]{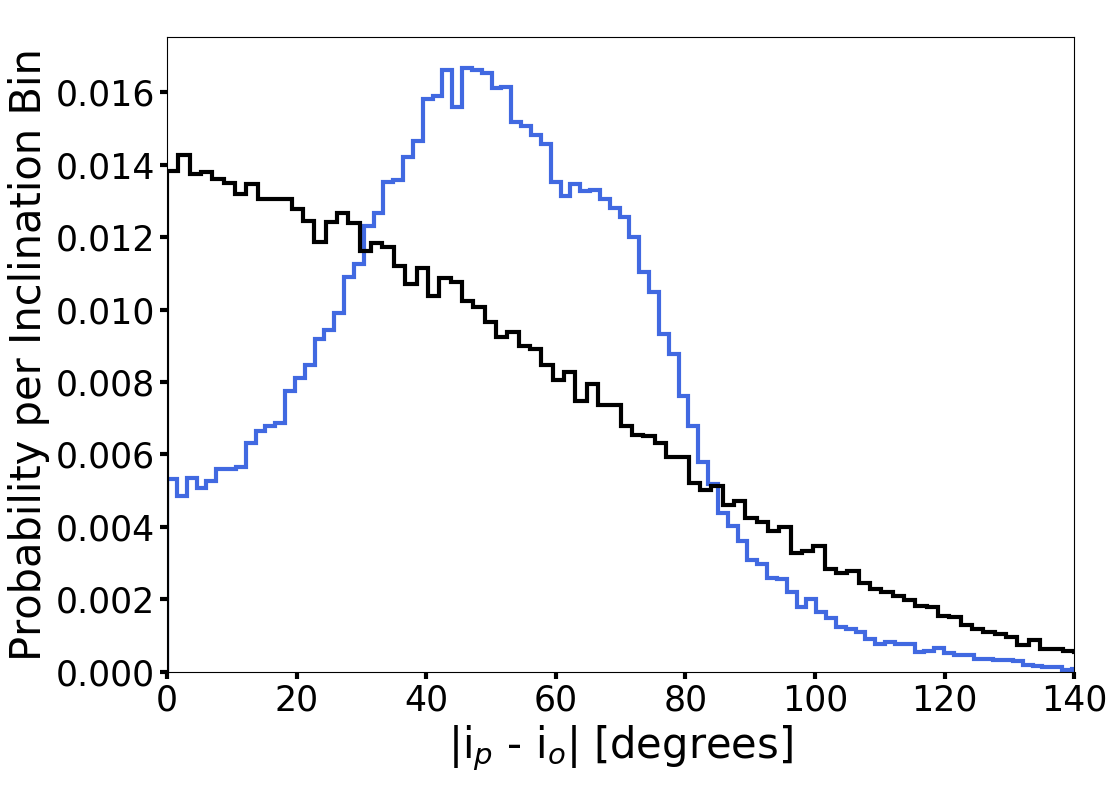}
\includegraphics[width=0.5\textwidth]{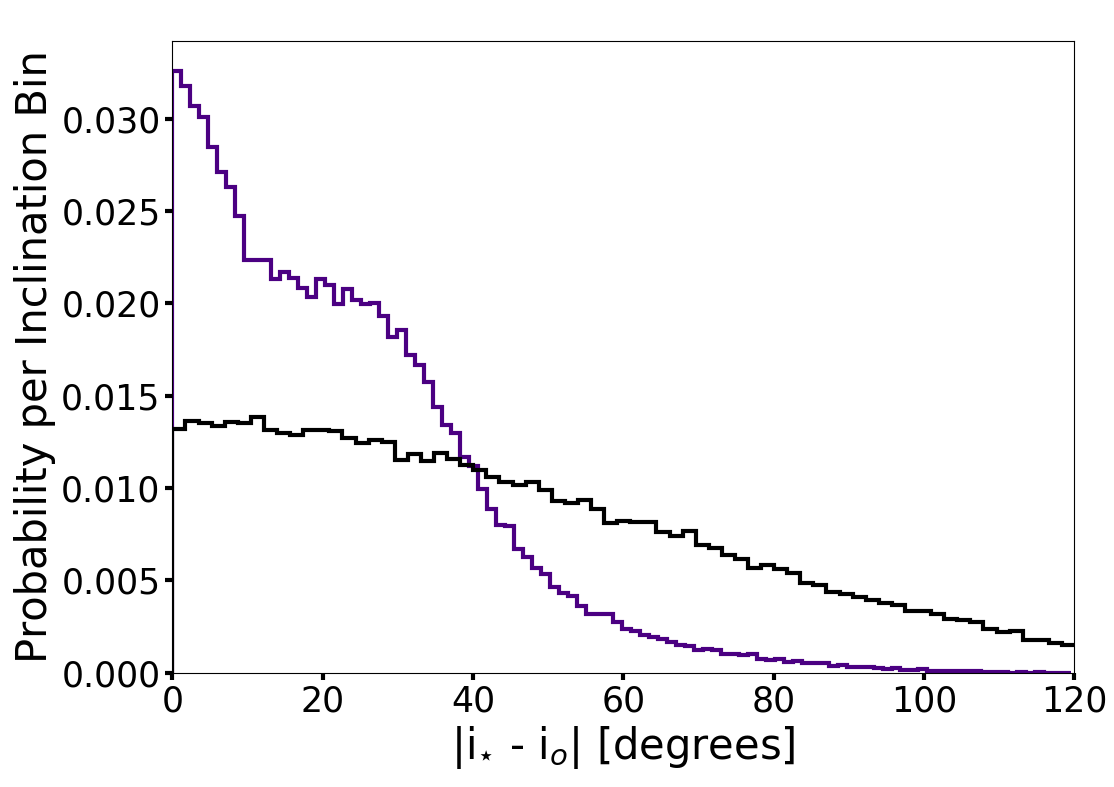}
\includegraphics[width=0.5\textwidth]{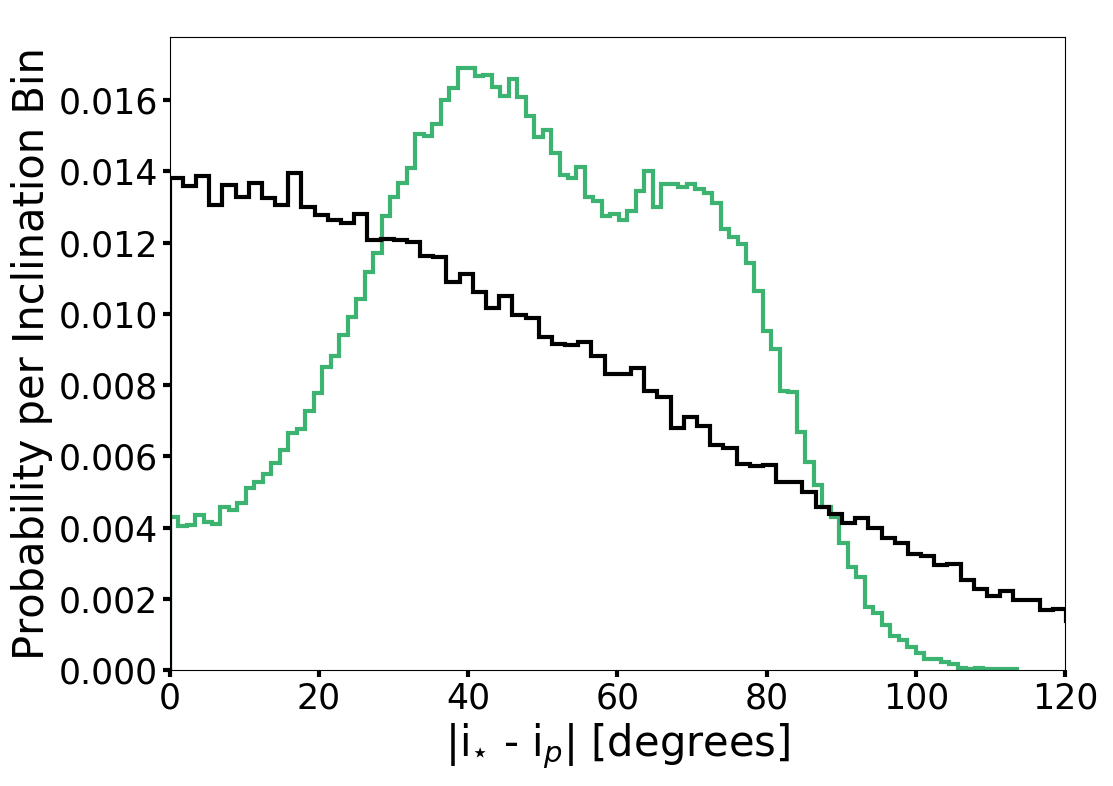}
\caption{Top panel: Posterior distribution of the line-of-sight companion obliquity (blue), whose value
is most probably $48^{+28}_{-21}$ degrees (all 
intervals are quoted at the 68\%
confidence interval).  Middle panel: Posterior distribution of the line-of-sight stellar obliquity (purple), whose
value is most probably $1^{+28}_{-1}$ degrees.  Bottom panel:  Posterior distribution of the relative inclination between line-of-sight stellar and PMC spin axes (green), most probably equal to $44^{+32}_{-16}$ degrees.  In each panel we show a random obliquity distribution for comparison (black).
Note that all of these line-of-sight mutual inclinations
are lower limits to the true de-projected mutual inclinations.}
\label{fig: obliquity posteriors}
\end{figure}

Ultimately we want to measure true de-projected companion and stellar obliquities $\Psi_p$ and $\Psi_{\star}$: 
\begin{equation} \label{true_obl}
\Psi_p = \cos^{-1}(\cos i_p\cos i_o + \sin i_p \sin i_o \cos \lambda_p)
\end{equation}
\begin{equation} \label{true_obl_star}
\Psi_{\star} = \cos^{-1}(\cos i_{\star}\cos i_o + \sin i_{\star} \sin i_o \cos \lambda_{\star}) 
\end{equation}
\noindent where $\lambda_p$ is the longitude of
ascending node of the companion's spin equatorial plane on its
orbital plane (also known as the sky-projected spin-orbit angle)
and $\lambda_\star$ is the analogous angle for the stellar spin. Neither $\lambda_p$ nor $\lambda_\star$ is known. 
Nevertheless, from the above equations we see that 
the absolute difference between the line-of-sight spin axis inclination and orbital inclination yields a lower limit on the true de-projected obliquity (i.e., the value of $\Psi$ when $\lambda = 0$; e.g., \citealt{Bowler2017}):
\begin{align} \label{obl_limit}
\Psi_p &> |i_p - i_o| \\
\Psi_\star &> |i_\star - i_o| \,.
\end{align}
In Figure \ref{fig: obliquity posteriors}, we show the probability distributions for 
$|i_p - i_o|$ (a.k.a.~the line-of-sight companion obliquity),
$|i_{\star} - i_o|$ (the line-of-sight stellar obliquity), 
and for completeness $|i_{\star} - i_p|$. Each of these is compared
against random distributions of line-of-sight inclination differences computed
by drawing $i_\star$, $i_p$, and $i_o$ from distributions uniform in $\cos i$.

We find that the line-of-sight companion obliquity
$|i_p - i_o|$ = $48^{+28}_{-21}$ degrees---this inclination difference prefers to be large, more so than in a random distribution,
as can be seen from Figure \ref{fig: obliquity posteriors}.
By contrast, we find that the line-of-sight stellar obliquity $|i_{\star} - i_o|$ = 1$^{+28}_{-1}$ degrees---this inclination difference prefers
to be small, more so than in a random distribution. 
Finally, the relative angle between line-of-sight companion and stellar spin axes 
is $|i_{\star} - i_p|$ = $44^{+32}_{-16}$ degrees;
our data favor greater misalignment between these spin vectors than if they were distributed isotropically.

\begin{figure}[h]
\centering
\includegraphics[width=0.5\textwidth]{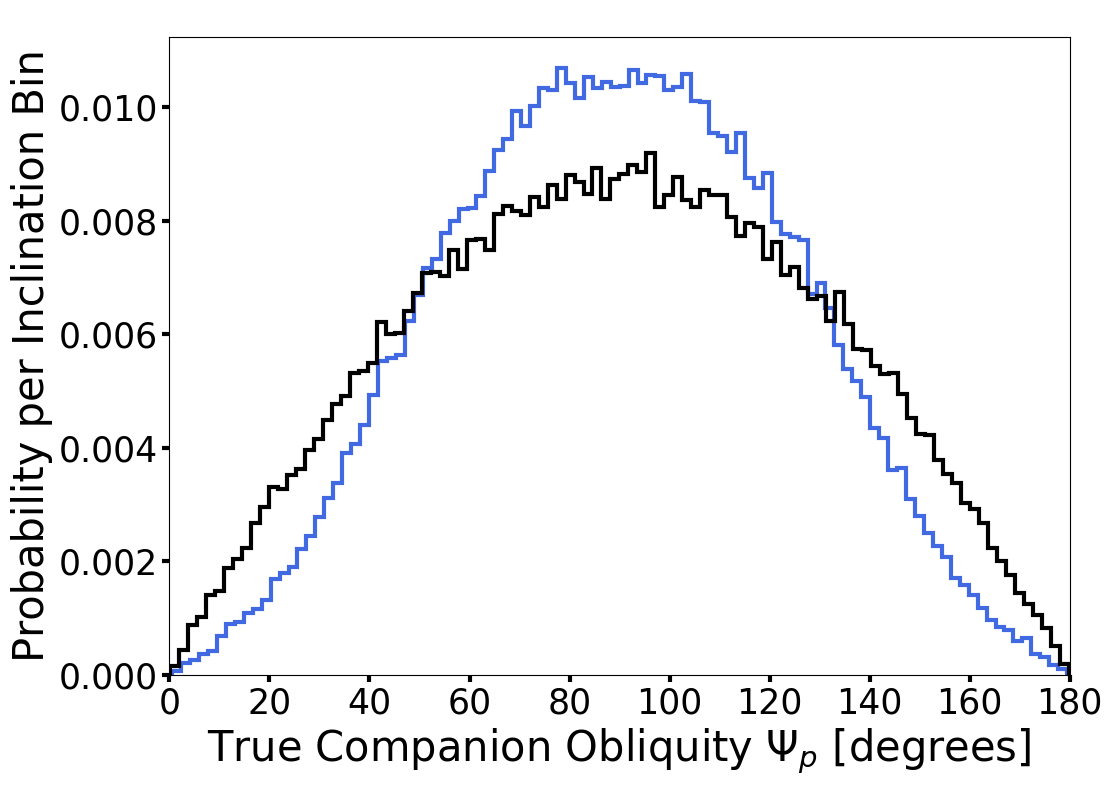}
\caption{Normalized posterior distribution of the true de-projected PMC obliquity $\Psi_p$ (blue), compared with a random distribution (black). The data marginally prefer
companion obliquities that are more misaligned.}
\label{fig: companion obliquity posterior}
\end{figure}

\begin{figure}[h]
\centering
\includegraphics[width=0.5\textwidth]{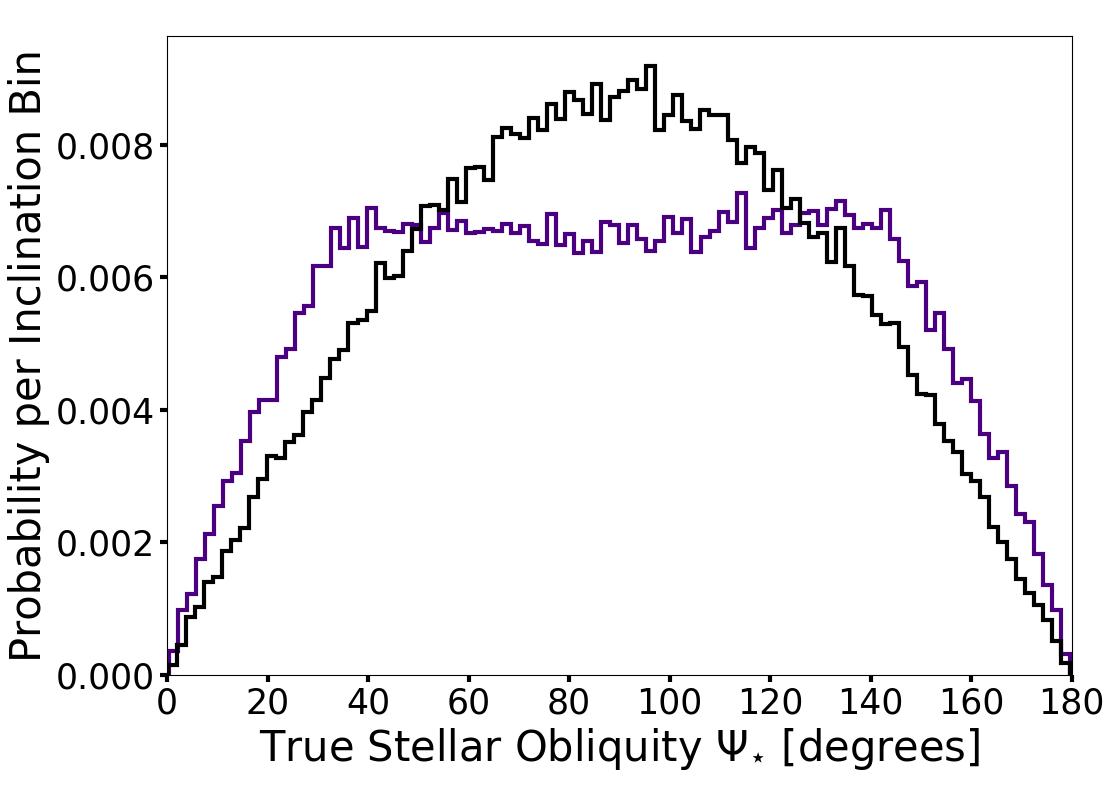}
\caption{Normalized posterior distribution of the true de-projected stellar obliquity $\Psi_\star$ (purple), compared
with a random distribution (black). The former
tentatively favor more aligned stellar obliquities as compared
with the latter.}
\label{fig: stellar obliquity posterior}
\end{figure}

Taking stock of our findings so far: we have found evidence
that the lower limit on $\Psi_\star$ is small, 
and that the lower limit on $\Psi_p$ 
is large. 
We can also compute probability distributions for $\Psi_\star$ and $\Psi_p$ directly using equations (\ref{true_obl}) and (\ref{true_obl_star}),  
assuming {\it a priori} uniform distributions for $\lambda_\star$ and $\lambda_p$, respectively. 
The resulting posteriors are shown in Figures
\ref{fig: companion obliquity posterior}
and \ref{fig: stellar obliquity posterior}, together with their random counterparts. 
Clearly not knowing $\lambda_p$ and $\lambda_\star$
opens up
a wide range of possibilities for $\Psi_p$
and $\Psi_\star$, and our posteriors
for these obliquities are roughly similar to random
distributions; in particular they all have modes
near $90^\circ$ as there is simply more phase space
there (it is for the same reason that edge-on orbits
are more commonplace than face-on orbits).

Nevertheless Figures \ref{fig: companion obliquity posterior}
and \ref{fig: stellar obliquity posterior} also
reveal notable differences
between our posteriors and the random distributions.
We may ask, given our data, how probable is it that the true companion obliquity is misaligned rather than aligned?  We calculate a Bayesian odds ratio to answer this question, defining ``aligned'' obliquities to be between 0 and 20 degrees, and ``misaligned'' obliquities to be between 20 and 180 degrees.  The probability of misalignment given our measurements is determined by integrating the posterior obliquity distribution from 20 - 180 degrees, while the prior on a misaligned obliquity is found by integrating the random obliquity distribution over the same range of angles.  Aligned probabilities are calculated in the same fashion, replacing the integration ranges with 0 - 20 degrees.  The odds ratio of misaligned-to-aligned true companion obliquities is 2.4:1; companion misalignment is preferred at 1.1$\sigma$ significance.  The same calculation for $\Psi_\star$ yields an odds ratio of 1:1.6; stellar alignment is preferred at 0.9$\sigma$ significance.

\begin{deluxetable*}{ccc}
\tablecaption{ Measured Parameters \label{tb:obl}}
\tabletypesize{\footnotesize}
\tablehead{
  \colhead{Parameter} & 
  \colhead{Measured Value} & 
  \colhead{Ref}
}
\startdata
$v_p \sin i_{p}$ & $13.4^{+1.4}_{-1.2}$ km/s & This work \\
$v_\star \sin i_{\star}$ & $18.2^{+0.5}_{-0.4}$ km/s & This work \\
$P_{rot,p}$ & 6.0$^{+2.6}_{-1.0}$ hrs & \citet{Zhou2019} \\
$P_{rot,\star}$ & 1.49 $\pm$ 0.02 days & This work \\
Astrometry & see Table 1 & This work; \citet{Bowler2013,Bryan2016}\\
$i_p$ & 33$^{+17}_{-9}$ or 147$^{+9}_{-17}$ deg & This work \\
$i_{\star}$ & $75 \pm 8$ or $105 \pm 8$ deg & This work \\
$i_o$ & 103$^{+16}_{-6}$ deg & This work \\
$|i_o - i_p|$ & $48^{+28}_{-21}$ deg & This work \\
$|i_o - i_{\star}|$ & 1$^{+28}_{-1}$ deg & This work \\
$|i_{\star} - i_p|$ & $44^{+32}_{-16}$ deg & This work 
\enddata
\tablecomments{The angles $i_p$ and $i_{\star}$ are both symmetric about 90 degrees due to the fact that we do not know whether these spin angular momentum vectors are pointing towards us or away from us. The angles presented here are all line-of-sight inclinations.  As described in section 3.7, the line-of-sight obliquities $|i_o - i_p|$ and $|i_o - i_{\star}$ are lower limits on the true de-projected obliquities $\Psi_p$ and $\Psi_{\star}$.}
\end{deluxetable*}

\section{Discussion: Possible Formation Histories}

We assess possible origin scenarios
for the 2M0122 system. In section \ref{3D} we presented evidence that
the obliquity of the companion
2M0122b is large, and that the obliquity
of the host star 2M0122 is small.
That evidence is marginal at present.
Nonetheless, to focus our discussion,
we assume a large companion obliquity
and a small stellar obliquity
and ask what dynamical histories are compatible.

\subsection{Physical collision} \label{physcol}
Might 2M0122b have suffered a collision that knocked it on its
side, as has been speculated for Uranus \citep[e.g.][]{Kegerreis2018}?
We think that scenario is unlikely. The escape velocity
from the surface of 2M0122b is
\begin{align}
v_{\rm esc,p} &= \sqrt{2GM_{\rm p}/R_{\rm p}} \nonumber \\
&\simeq 270 (M_{\rm p}/20 M_{\rm J})^{1/2} (R_{\rm J}/R_{\rm p})^{1/2} \, {\rm km}/{\rm s} \nonumber
\end{align}
where $M_{\rm p}$ and $R_{\rm p}$ are the companion mass
and radius, respectively, and $G$ is the gravitational constant. This is
considerably larger than the escape velocity
from the star at the orbital semimajor axis
of 2M0122b,
\begin{align}
v_{\rm esc,\star} &= \sqrt{2GM_\star/a_{\rm p}} \nonumber \\
&\simeq 4 (M_\star/0.4M_\odot)^{1/2} (50\, {\rm AU}/a_{\rm p})^{1/2} \, {\rm km}/{\rm s} \nonumber
\end{align}
where $M_\star$ is the host stellar mass and $a_{\rm p}$
is the orbital distance.
Since the ratio (a.k.a.~the square root of the Safronov number)
\begin{equation}
v_{\rm esc,p}/v_{\rm esc,\star} \sim 70 \gg 1 \,,
\end{equation}
2M0122b is more likely to have ejected a neighboring body out of the system than to have collided with it; i.e., unless the geometry of the encounter
were fine-tuned, any object interacting gravitationally with 2M0122b would have its velocity
excited to $> v_{\rm esc,\star}$ and be ejected 
before it could physically
collide. (We note in passing that
$v_{\rm esc,p}/v_{\rm esc,\star}$ evaluates to 2 for Uranus.)

\subsection{Secular spin-orbit resonance induced by a perturber}\label{sores}
A secular spin-orbit resonance is a commensurability between the frequencies of the planet's spin axis precession and its orbital precession, the latter of which may arise from interactions with an as-yet-undetected planetary perturber. Secular spin-orbit resonances can excite planetary obliquities to large values during the system's formation. These resonances are common in the Solar System. For instance, Saturn's $27^{\circ}$ obliquity may be due to a secular spin-orbit resonance with Neptune \citep{2004AJ....128.2501W, 2004AJ....128.2510H}. The obliquities of Jupiter \citep{2006ApJ...640L..91W} and possibly Uranus and Neptune \citep{2019arXiv190810969R} are also thought to be affected by this mechanism. Moreover, all of the terrestrial planets likely experienced chaotic obliquity variations in their past due to overlap of multiple secular spin-orbit resonances \citep{1993Natur.361..608L}. Mars' obliquity is still in a chaotic state for this reason \citep{1993Sci...259.1294T}. 

Among extrasolar planets, secular spin-orbit resonances are also thought to be common. \textit{Kepler} multi-planet systems -- which are composed of  short-period, compact, nearly-coplanar planets within $P \lesssim 100$ days -- are in a regime of parameter space that makes them intrinsically susceptible to these resonances, suggesting that they may frequently have large obliquities \citep{2019NatAs...3..424M}. Secular spin-orbit resonances have not yet been investigated for exoplanets with semi-major axes as large as several tens of AU. Here we examine whether this resonance is possible for 2M0122b.

We start by defining the frequencies of spin axis precession and orbital precession.
The torque from the host star on the rotationally-flattened figure of 2M0122b causes the planet's spin axis to precess about its orbital angular momentum vector. The period of this precession is
\begin{equation} \label{talpha}
T_{\alpha}=2\pi/(\alpha\cos\Psi_p), 
\end{equation}
where $\alpha$ is the spin axis precession constant and $\Psi_p$ is the companion obliquity. In the absence of satellites orbiting 2M0122b, $\alpha = \alpha_0$ is given by \citep{1997A&A...318..975N, 2003Icar..163....1C} 
\begin{equation}
\alpha_0=\frac{1}{2}\frac{M_{\star}}{M_p}\left(\frac{R_p}{a}\right)^3\frac{k_{2,p}}{C_p}\frac{\omega_p}{(1-e^2)^{3/2}}.
\label{alpha}
\end{equation}
Here $k_{2,p}$ is the planet's Love number, a dimensionless value related to the planet's central concentration and its deformation response to tidal disturbance. The quantity $C_p$ is the planet's moment of inertia normalized by $M_p {R_p}^2$. Finally, $\omega_p = 2\pi/{P_{rot,p}}$ is the spin angular frequency.

If 2M0122b is accompanied by one or more satellites, its spin axis precession frequency may be enhanced as a result of the 
adiabatic gravitational coupling between the satellite(s) and the oblate planet \citep{1965AJ.....70....5G}. We define $f_{\alpha} = \alpha/\alpha_0$ as the enhancement factor of the spin-axis precession frequency over the satellite-free case. 
For a satellite having a mass ratio with respect to the planet of $m_s/M_p = 10^{-3}$ and 
occupying a circular orbit 
at the greatest possible separation for adiabatic gravitational coupling, $a_s/R_p \sim 220$, the maximum value of the frequency enhancement is $f_{\alpha} \sim 10^4$ 
(\citealt{2019ApJ...876..119M}, see their Figure 2).

In addition to the spin axis precession frequency, the other relevant frequency for secular spin-orbit resonance is the orbit nodal recession frequency. If there is another planet in the 2M0122 system, secular planet-planet interactions will drive nodal recession with a frequency, $g = |\dot{\Omega}|$, where $\Omega$ is the longitude of the ascending node. This frequency may be calculated using Laplace-Lagrange theory if the eccentricities and mutual inclination are small. This is not adequate in our context given the preference for high eccentricity orbits shown in Section \ref{sec:measuring i_orbit}. We adopt the hierarchical three-body secular approximation and use the octupole-order expansion from \cite{2013MNRAS.431.2155N}. The equations of motion for $\dot{\Omega}$ may be found in their Appendix B, or alternatively, in \cite{2016ARA&A..54..441N}, and it is suitable for both the interior perturber and exterior perturber cases.

A secular spin-orbit resonance is an instance of a Cassini state, an equilibrium configuration of the planet's spin vector in a uniformly precessing orbit frame \citep{1966AJ.....71..891C, 1969AJ.....74..483P}. Up to four Cassini states may exist depending on the ratio of $g/{\alpha}$; Cassini state 2 is the most favorable for obtaining a large obliquity with prograde rotation. Cassini states obey the equilibrium relation
\begin{equation}
\label{Cassini state criterion}
g\sin(\Psi_p - I) - \alpha\cos\Psi_p \sin \Psi_p = 0
\end{equation}
where 
$I$ is the orbital inclination with respect to the invariable plane.

To determine the possibility of secular spin-orbit resonance for 2M0122b, we apply a Monte Carlo procedure that explores the  allowable parameter space. We begin by sampling all system parameters according to their observational constraints or theoretical allowances. We sampled 2M0122b's semi-major axis, $a_b$, and eccentricity, $e_b$, from the posterior obtained in the orbit fit from Section \ref{sec:measuring i_orbit}. The remainder of the parameters are listed in Table \ref{Secular spin-orbit resonance parameter table}. The parameter $i_{\mathrm{mut}}$ is the mutual orbital inclination between 2M0122b and the perturber. For each sample, we calculated the obliquity that 2M0122b would require to satisfy criterion (\ref{Cassini state criterion}) and be in Cassini state 2. Finally, we discarded samples for which the calculated obliquity was not in the observed range.

\begin{table}[h]
\centering
\caption{Parameters and their ranges used for the Monte Carlo investigation of secular spin-orbit resonance.}
\begin{tabular}{c | c}
\hline
\hline
\multicolumn{2}{c}{2M0122b parameters} \\
\hline
Parameter & Range \\
\hline
$M_b$ ($M_{\mathrm{J}}$) & (12, 14), (23, 27)    \\
$R_b$ ($R_{\mathrm{J}}$) & (0.86, 1.22) \\ 
$P_{{\mathrm{rot}, p}}$ (hr) & (5.0, 8.6) \\ 
$k_{2,p}$ & (0.05, 0.6) \\
$C_p$ & (0.1, 0.3) \\
$f_{\alpha}$ & (1, $10^4$) \\
$a_b$ (AU) & MCMC from \S \ref{sec:measuring i_orbit} \\
$e_b$ & MCMC from \S \ref{sec:measuring i_orbit} \\
\hline
\hline
\multicolumn{2}{c}{Perturber parameters} \\
\hline
Parameter & Range \\
\hline
$M_c$ ($M_{\mathrm{J}}$) & (0.1, 7)    \\
$a_c$ (AU) & (1, 400) \\
$e_c$ & (0, 1) \\
$i_{\mathrm{mut}}$ & ($0^{\circ}$, $90^{\circ}$) \\ 
\end{tabular}
\label{Secular spin-orbit resonance parameter table}
\end{table}

\begin{figure}[h]
\centering
\includegraphics[width=0.5\textwidth]{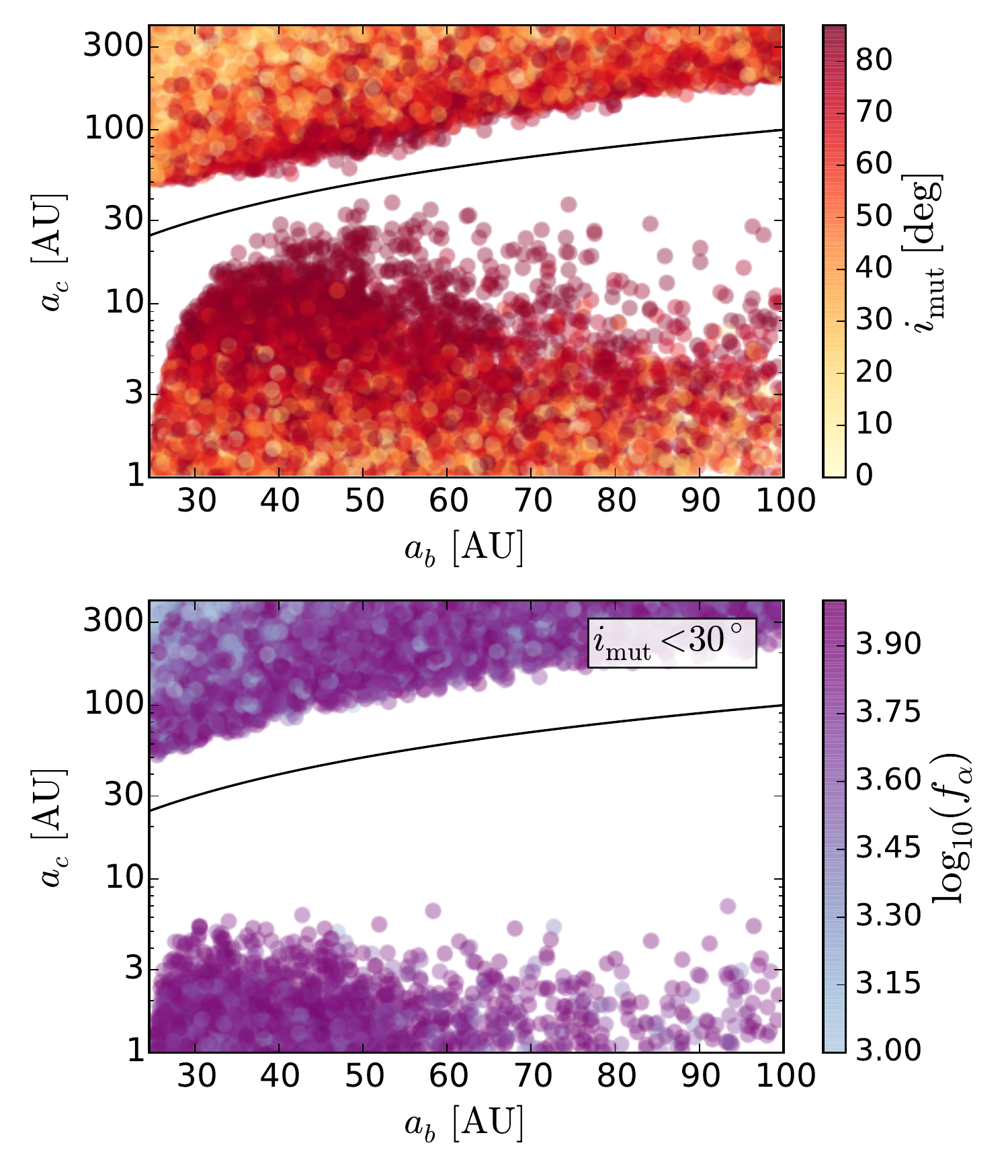}
\caption{ Results of the Monte Carlo investigation of a secular spin-orbit resonance for 2M0122b. The plotted points are samples in $a_b$ -- $a_c$ space for which the planet obliquity satisfying Cassini state 2 was within the range obtained from our measurements. \textit{Top:} The colorbar is the mutual orbital inclination, $i_{\mathrm{mut}}$. \textit{Bottom:} Restricted to samples with $i_{\mathrm{mut}} < 30^{\circ}$. The colorbar indicates $f_{\alpha}$, the satellite-induced enhancement of $\alpha$. The solid black line in both panels is $a_b = a_c$, dividing the domain between interior and exterior perturbers.}
\label{fig: secular spin-orbit resonance Monte Carlo}
\end{figure}

The results of this Monte Carlo investigation are shown in Figure \ref{fig: secular spin-orbit resonance Monte Carlo}. In both panels, the plotted points are samples in $a_b$ -- $a_c$ space. We see that the perturber would need to be 
fairly distant from 2M0122b, $a_c \lesssim 10$ AU or $a_c \gtrsim 100$ AU. Simultaneously, 2M0122b must be orbited by a satellite that is large enough to substantially enhance the spin-axis precession frequency. This can be seen in the bottom panel with $i_{\mathrm{mut}} < 30^{\circ}$, where the colorbar indicates that the frequency enhancement must be $f_{\alpha} \sim 10^4$. These restrictive requirements are due to 2M0122b's distant orbit. The semi-major axis is so large that the spin-axis precession period is naturally very long ($\alpha$ very small). The resonant commensurability $g\sim\alpha$ is thus only possible through wide perturber separations (which decreases $g$) and the presence of a satellite (which increases $\alpha$). 

Given the extreme and fine-tuned nature of these parameter constraints, it 
appears unlikely that a secular spin-orbit resonance is the cause of 2M0122b's obliquity excitation.

\subsection{Kozai-Lidov oscillations from an external perturber}\label{KL oscillations}

The Kozai-Lidov (KL) effect involves orbital inclination and eccentricity oscillations driven by an external perturber.
The mechanism can also excite planetary obliquities. 
The orbital inclination oscillations induced by the perturber will inevitably produce large values of the planetary obliquity as long as the KL oscillation period is shorter than the planet's spin precession period $T_\alpha$ (equation \ref{talpha}).
Applying KL oscillations to the 2M0122 system could explain both the observed non-zero obliquity and the high orbital eccentricity. 
This picture would further predict
that the stellar obliquity is large.
It has been shown that a planet undergoing KL oscillations may induce chaotic variations in the stellar spin axis,
with vanishingly small probability of observing
the stellar spin aligned with the orbit normal 
(e.g., \citealt{2014Sci...345.1317S}, their Figure 1, bottom panel). While a large
stellar obliquity is allowed by our
data (Figure \ref{fig: stellar obliquity posterior}), alignment is tentatively preferred at the 0.9$\sigma$ level
(Figure \ref{fig: obliquity posteriors}).

Perhaps the primary objection 
to this scenario is the obvious one:
the lack of an observed 
external perturber to 2M0122b.
Deep imaging of the system 
rules out a companion more massive than 7 M$_{\rm J}$ between 30-200 AU \citep{Bryan2016}. 
This upper limit is constraining,
as KL oscillations 
require an external companion
having more orbital angular momentum
than 2M0122b, which has a mass
$\sim$12-27  M$_{\rm J}$
and resides at an orbital
distance of $\sim$50 AU.

\subsection{``Twisting" the orbit but not the spin}\label{twist}
The Kozai-Lidov mechanism considered in \S\ref{KL oscillations}
is just one way in which an
external torque can ``twist'' (change the direction of)
a planet's orbital angular momentum vector \citep{Tremaine1991}.
For a twist of whatever cause to also change the obliquity,
it must occur quickly, on a timescale shorter than
the planet's spin precession period $T_\alpha$ (equation \ref{talpha});
otherwise, for slow twists, 
the spin vector adiabatically tracks the
orbit normal and the obliquity does not change.

The most favorable case for obliquity changes would seem to be when 
2M0122b's spin precession is controlled only by the host stellar torque;
then $T_\alpha \sim 10^{11}$ yr, long enough that any external agent acting within the 
$10^8$ yr system age could provide a fast enough twist. 
However, if we believe that
the stellar obliquity is small (and again, the evidence for this
is marginal), then we must select for those processes that do not alter it from its
presumed primordially small value.
Continuing to assume that 2M0122b is the only companion to the star,
we find the stellar spin-orbit coupling to be even weaker
than the companion spin-orbit coupling; the stellar spin axis 
precesses about the orbit normal over
$T_{\alpha,\star} \sim 10^{12}$ yr. 
It would seem that no twist can both excite the companion obliquity and preserve the stellar obliquity
while the 2M0122 system is in its current configuration.

\citet{Tremaine1991} arrived at an analogous conclusion for the Solar
System. He argued that any twist scenario to explain the obliquities of the outer planets while also keeping the Solar obliquity at its present
modest (7$^\circ$) value cannot be staged after the Solar System's formation but must take place during it, while infall from the parent molecular cloud and disk accretion are still ongoing.
The external torque that tilts a planet's orbit
may be exerted by ``inhomogeneities'' in the cloud or disk; these same ``mass concentrations,'' carrying
anomalous angular momentum, may ultimately be accreted by the proto-Sun
and help to re-align its spin axis with the disk axis. Re-alignment would also be effected by gravitational forces between the disk and young Sun, which owing to its faster primordial spin
would have a larger rotational bulge for stronger 
gravitational coupling.

The large obliquity of 2M0122b also
seems most easily understood in the context of its
formation within an accreting circumstellar disk.
This idea is supported by
the considerations of this subsection, and of
preceding subsections \S\ref{physcol}--\ref{KL oscillations},
all of which
point to difficulties in generating the large
obliquity after the companion's formation, in a disk-less
setting. We turn now to what that formation environment
might look like.

\subsection{Formation within a gravito-turbulent disk}\label{gravturb}
\citet{Nielsen2019} found evidence that the demographics of objects with masses $> 12 M_{\rm J}$ (``brown dwarfs'') are distinct from those of less massive objects (``giant planets'').  Specifically, brown dwarfs exhibit a top-heavy mass function, an orbital distance distribution weighted toward large separations ($\sim$100 AU), and no preference for host stellar spectral type---and in particular no preference for massive host stars. All these trends reverse for objects less massive than $12 M_{\rm J}$.  While their study lacked the statistics to determine with confidence the exact dividing mass between brown dwarfs and giant planets, it appears increasingly clear that high-mass PMCs and low-mass PMCs form differently. \citet{Nielsen2019} described how the demographic trends exhibited by brown dwarfs are correctly predicted by top-down formation by gravitational instability, while those of giant planets are consistent with bottom-up formation by core accretion (their section 6.3; see also \citet{Wagner2019}).

On the face of it, with a mass $> 12 M_{\rm J}$, an orbital separation of $\sim$50 AU, and a low-mass stellar host, the PMC 2M0122b possesses all the properties of a brown dwarf as defined and characterized by \citet{Nielsen2019}. We consider the possibility that 2M0122b formed by gravitational instability, and ask whether such a scenario can accommodate large companion obliquities.

The criteria for gravitational collapse (a Toomre $Q$ parameter $\lesssim 2$, and a disk cooling time shorter than the orbital time; \citep{Gammie2001}) are typically satisfied at large stellocentric distances in young disks still being fed by their natal clouds \citep{Kratter2016}. These are dynamically active environments; in addition to continued infall, overdensities in the disk spontaneously form and shear away.  Fluid random motions in ``gravito-turbulent'' disks are vigorous---they are trans-sonic when cooling times approach orbital times, and fully three-dimensional.  See Figure 7 and in particular Figure 9 of \citet{Shi2014}, which shows a meridional flow where gas that is compressed radially from self-gravity is directed vertically out of the midplane, and accelerated back down.

An overdensity that cools fast enough to break off from the background turbulence and become self-bound---the clump that in this scenario eventually becomes 2M0122b---will be gravitationally torqued by surrounding overdensities, both unbound and bound. Overdensities off the disk midplane may not only twist a bound clump's orbit (\S\ref{twist}) but may also directly change the clump's spin vector by gravitationally coupling to its rotational bulge. That bulge may take the form of a fully rotationally supported, ``circumplanetary disk'' (CPD) on the scale of the proto-brown dwarf's Hill or Bondi radius. The CPD may ``wobble'' in response to stochastic gravitational forcing. Direct accretion of gas from the circumstellar disk onto the proto-brown dwarf will also change its obliquity, as the gravito-turbulent gas will arrive onto the clump from a variety of angles---probably mostly from out-of-midplane directions \citep[e.g.][]{Fung2019}.

Disk overdensities, including flocculent spiral arms that come and go, are also expected to gravitationally perturb the nascent PMC onto an eccentric orbit.  This is qualitatively consistent with the observational evidence presented in Section \ref{sec:measuring i_orbit} for 2M0122b having a large orbital eccentricity. The eccentricity may also grow via interactions between the PMC and  disk at the outer 1:3 Lindblad resonance, a mechanism shown to apply
only to objects $\gtrsim 5$--$20 M_{\rm J}$ orbiting Sun-like stars
\citep{Papaloizou2001, Kley2006, Dunhill2013, Bitsch2013}.

The overall picture is of the young 2M0122b immersed in a fully 3D disk filled with strong density fluctuations which randomly force its orbit and obliquity. The star's obliquity is
presumably maintained at a primordially small value, as the disk in the vicinity
of the star has a scale height smaller than the stellar radius and 
therefore behaves as a 2D sheet; any misalignment between
the disk and the stellar rotational bulge would be zeroed out
by their mutual gravity.

This picture needs to be fleshed out quantitatively.
In addition to exerting stochastic gravitational torques on the CPD, the circumstellar disk also exerts a mean gravitational torque. The mean torque may
bring the CPD into alignment with itself and zero out the companion
obliquity, although interestingly there is another possibility:
if the CPD is sufficiently inclined to the circumstellar disk,
the inclination (obliquity) may be driven to Kozai's critical angle of 39$^\circ$ \citep[e.g.][]{Martin2014}.
Magnetic fields generated within 2M0122b may also play a role. A case has been made that magnetic coupling of a PMC's magnetosphere to its CPD is essential for explaining the spin periods of both brown dwarfs and giant planets, observed to be 5--20\% of break-up \citep{Bryan2018,Batygin2018,Ginzburg2019}. The spin period of 2M0122b falls squarely in this range and presumably reflects the same magnetospheric regulation. Again, to explain
a large obliquity, the CPD
would have to be tilted out of the circumstellar disk plane.

\section{Conclusions}

In this study we constrained for the first time all three angular momentum vectors in a substellar system. The host star, 2M0122, has
a mass of 0.4 M$_{\odot}$ and hosts a directly imaged 12-27 M$_{\rm J}$ companion at
50 AU. 
We measured how the stellar spin angular momentum vector, the companion spin angular momentum vector, and the orbital angular momentum vector are inclined relative to the sky plane (Fig.~1).
Underlying these measurements are five direct observables: projected rotation rates ($v\sin i$) for both companion and star, rotation periods ($P_{\rm rot}$) for both companion and star, and the astrometric orbit of the companion.  

The projected rotation speeds 
for the star and companion were obtained from near-infrared high-resolution NIRSPEC/Keck spectra. To the published photometric rotation period for the companion 2M0122b measured using \textit{HST} \citep{Zhou2019} we added the photometric rotation period for the star using \textit{TESS}. 
We found that while the stellar spin axis is nearly perpendicular to our line of sight---$i_\star = 75 \pm 8$ degrees---the companion spin axis is decidedly not: $i_p = 33^{+17}_{-9}$ degrees. Note that the posterior probability distributions for $i_p$ and $i_{\star}$ are symmetric about 90 degrees (Figures \ref{fig: planetary spin axis posterior} and \ref{fig: stellar spin axis posterior}).

We fitted nine epochs of astrometry, including one new epoch from NIRC2/Keck that doubled the astrometric baseline.  
The orbital inclination $i_o$ relative to the sky plane
is $103^{+16}_{-6}$ degrees (Figure \ref{fig: orbital inclination posterior}):
we are viewing the orbit nearly
edge on.

From the posterior probability
distributions of $i_o$, $i_p$, and $i_\star$
we computed posteriors for 
$|i_p - i_o|$ and $|i_{\star} - i_o|$. 
These ``line-of-sight obliquities''
for companion and star are lower limits
on the true de-projected obliquities
$\Psi_p$ and $\Psi_\star$.
We found that while the line-of-sight stellar obliquity is
small, $|i_\star - i_o|$ = $1^{+28}_{-1}$ degrees, the line-of-sight companion obliquity is large, $|i_p-i_o| = 48^{+28}_{-21}$ degrees.
Moreover, these preferences for a small
lower bound on the stellar obliquity and a large lower
bound on the companion obliquity are each stronger
than for a random distribution of spin and orbit
vectors (Figure \ref{fig: obliquity posteriors}).

We also computed posteriors for the true 3D obliquities
$\Psi_p$ and $\Psi_\star$ by assuming a uniform
prior on the unknown spin-orbit angle $\lambda$ (the
node of the spin equatorial plane on the orbit plane).
Although these posteriors admit
a wide range of values,
they still deviate from a purely random distribution in 
suggestive ways: compared to a (data-free) distribution
that is uniform in $\cos \Psi$, our posteriors
tentatively favor spin-orbit alignment for the
star at the 0.9$\sigma$ level and spin-orbit misalignment for the companion at the 1.1$\sigma$ level.

For the parameters of the 2M0122 system,
a large companion obliquity and a small stellar
obliquity are perhaps most readily understood 
if the companion formed by disk gravitational instability.
Other possibilities for explaining such
obliquities---collisions, secular spin-orbit
resonance, and Kozai-Lidov oscillations---are disfavored.
In a 3D gravito-turbulent circumstellar disk, 
fragments that break off and become self-bound
should have a wide dispersion of spin vectors,
reflecting the strongly turbulent velocity field
from which they were drawn and interact,
either by accretion or stochastic gravitational
forcing. Note that formation of 2M0122b
by gravitational instability
is favored on independent grounds,
as this PMC shares the same properties
of brown dwarfs as measured by \citet{Nielsen2019},
who explained how brown dwarf demographics 
could be reproduced by top-down gravitational collapse.
Because these multiple lines of evidence point to formation via gravitational instability, we believe 2M0122b is more appropriately classified a ``brown dwarf'' rather than a ``planet''.  Future theoretical work should examine the evolution
of fragment spins, including the orientation of
rotationally supported ``circumplanetary disks,''
and their magnetic regulation,
within self-gravitating circumstellar disks.

This work presents the first constraints on a planetary-mass companion obliquity 
outside the Solar System. As a new observable, obliquity 
presents an exciting and unique window into formation
history. The limitation of our study imposed
by sky projection (i.e., our ignorance of the spin-orbit
node $\lambda$) should be removed as more system obliquities
are measured---this will enable much more constraining
statistical studies.

\section*{}
We thank Ian Czekala, Courtney Dressing, Daniel Fabrycky, Jeffrey Fung, Sivan Ginzburg, and Yifan Zhou for helpful conversations.  M.L.B. is supported by her Heising-Simons Foundation 51 Pegasi b Fellowship.  E.C. acknowledges support from NASA and the National Science Foundation. B.P.B. acknowledges support from the National Science Foundation grant AST-1909209. S.M. is supported by the NSF Graduate Research Fellowship Program under Grant  DGE-1122492. S.B. is supported by a National Science Foundation Graduate Research Fellowship. This research made use of Lightkurve, a Python package for Kepler and TESS data analysis (Lightkurve Collaboration, 2018). This paper includes data collected with the \emph{TESS} mission, obtained from the MAST data archive at the Space Telescope Science Institute (STScI). Funding for the TESS mission is provided by the NASA Explorer Program. STScI is operated by the Association of Universities for Research in Astronomy, Inc., under NASA contract NAS 5-26555. We extend special thanks to those of Hawaiian ancestry on whose sacred mountain of Mauna Kea we are privileged to be guests. 

\bibliographystyle{aasjournal}
\bibliography{bibliography}

\end{document}